\documentstyle[psfig]{mn2e}

\begin{document}

\title[Comparison of 2dFGRS, SDSS and the MGC]{The Millennium Galaxy
Catalogue: The photometric accuracy, completeness and contamination 
of the 2dFGRS and SDSS-EDR \& DR1 datasets}
\author[Cross et al.]{\parbox[t]{\textwidth}
{N.J.G. Cross$^{1,*}$, S.P. Driver$^{2}$, J. Liske$^{3}$, D.J. Lemon$^4$, J.A. Peacock$^{5}$, S. Cole$^6$, P. Norberg$^{7}$, W.J. Sutherland$^{5}$}
\vspace*{6pt} \\ 
$^1$Department of Physics and Astronomy, John Hopkins University, Baltimore,
MD, 21218, USA \\
$^2$Research School of Astronomy \& Astrophysics, The Australian National 
University, Weston Creek, ACT 2611,Australia \\
$^{3}$ESO Headquarters Garching, Karl-Schwarzschild-Str. 2, D-85748 Garching bei M\"unchen, Germany \\
$^4$School of Physics and Astronomy, North Haugh, St Andrews, Fife, KY16 9SS \\
$^{5}$Institute for Astronomy, University of Edinburgh, Royal Observatory, 
Edinburgh EH9 3HJ \\
$^6$Department of Physics, South Road, Durham DH1 3LE \\
$^7$ETHZ Institut f\"ur Astronomie, HPF G3.1, ETH H\"onggerberg, CH-8093 Z\"urich, Switzerland \\
$^*${\rm (cross@pha.jhu.edu)}
}

\date{Accepted
...... Received .....}

\pagerange{\pageref{firstpage}--\pageref{lastpage}}


\label{firstpage}

\maketitle

\begin{abstract}
\noindent The Millennium Galaxy Catalogue (MGC) is a deep 
($\mu_{\rm B,lim}=26$ mag arcsec$^{-2}$), wide field CCD imaging survey, 
covering $37.5$deg$^2$. The MGC survey region is completely contained within 
the Two-degree Field 
Galaxy Redshift Survey (2dFGRS) and the Sloan Digital Sky Survey Early Data 
Release (SDSS-EDR). We compare the photometry and completeness of the 2dFGRS 
and the SDSS-EDR with the MGC over the range $16<B<20$ mag. We have also 
undertaken a photometric comparison to SuperCosmos and the SDSS First Data 
Release.

We find that $B_{\rm MGC}-B_{\rm 2dF}=(0.035\pm0.005)$ mag with an 
uncertainty of $0.142$ mag per galaxy, $B_{\rm MGC}-B_{\rm SCOS}=
(0.032\pm0.005)$ mag with an uncertainty of $0.108$ mag, 
$B_{\rm MGC}-B_{\rm SDSS-EDR}=(0.032\pm0.005)$ mag with an uncertainty 
of $0.094$ mag, and $B_{\rm MGC}-B_{\rm SDSS-DR1}=(0.039\pm0.005)$ mag with 
an uncertainty of $0.086$ mag. We find that high 
surface brightness 2dFGRS galaxies are systematically too faint, which leads 
to a significant scale error in magnitude. This effect is significantly 
reduced with the SCOS photometry. In the SDSS there is a weak non-linear 
scale error, which is negligible for faint galaxies. LSBGs in the SDSS are 
systematically fainter, consistent with the relative shallowness of this 
survey.

We find that the 2dFGRS catalogue has $(5.2\pm0.3)$ per cent stellar
contamination, $(7.0\pm0.4)$ per cent of objects resolved into 2 or more by
the MGC and is $(8.7\pm0.6)$ per cent incomplete compared to the MGC. From 
our all object spectroscopic survey we find that the
MGC is itself misclassifying  $(5.6\pm1.3)$ per cent of galaxies as stars,
hence the 2dFGRS misses $(14.3\pm1.4)$ per cent of the galaxy population. 
The SDSS-EDR galaxy catalogue
has $(1.3\pm0.1)$ per cent stellar contamination and $(5.3\pm1.0)$ per cent of
galaxies misclassified as stars, with $(0.18\pm0.04)$ per cent of objects
resolved into 2 or more by the MGC and is $(1.8\pm0.1)$ per cent incomplete
compared to the MGC. The total fraction of galaxies missing 
from the SDSS-EDR galaxy catalogue to $B_{\rm MGC} = 20$ mag, from 
incompleteness and misclassification is 
$(7.1\pm1.0)$ per cent.
\end{abstract}


\section{Introduction}
Galaxy surveys are now becoming sufficiently extensive that eyeball 
verification of the entire automated detection and classification 
algorithms are impractical. For example, the two-degree field 
Galaxy Redshift Survey input catalogue (Colless et al. 2001, 2003) contains 
over 300,000 galaxies for which redshifts have been targeted for 245,591
objects (229,118 galaxies, 16,348 stars and 125 QSOs) and high quality 
redshifts have been obtained for 221,414 galaxies.
Numerous publications, based on this dataset, have been used to constrain
cosmological parameters (e.g., Peacock et al. 2001; Efstathiou et al. 2002; 
Verde et al. 2002 and Elgaroy et al. 2002) and measure the local galaxy 
luminosity function(s) (see Folkes et al. 1999; Cole et al. 2001; 
Madgwick et al. 2002; Norberg et al. 2002b; de Propris et al. 2003), 
the bivariate brightness distribution (Cross et al. 2001; 
Cross \& Driver 2002), star-formation histories (Baldry et al. 2002; 
Lewis et al. 2002) and galaxy clustering (Norberg et al. 2001; 
Norberg et al. 2002a) for example. The credibility of these 
papers relies, to some extent, upon the underlying accuracy and uniformity 
of the photometric input catalogue and its completeness (see for example 
Colless et al. 2001, 2003, Norberg et al. 2002b, Cross et al. 2001 and 
Cole et al. 2001 which each discuss various aspects of completeness, 
reliability and various selection biases). 

Similarly the Sloan Digital Sky Survey which will eventually 
provide photometric information for over 1 billion objects (York et al. 2000) 
will also be publicly released in stages (e.g. Stoughton et al. 2002) and the 
value to the community will depend upon the accuracy of the automated 
photometric, astrometric and object classification algorithms (see for 
example Yasuda et al. 2001; Blanton et al. 2003). Here we aim to provide an 
independent estimate of the photometric and classification credibility 
of these public datasets, through the comparison with a third, manually 
verified dataset, namely the Millennium Galaxy Catalogue (MGC; Liske et al. 
2003; Lemon 2003). 

The MGC is particularly suitable as it covers a sufficiently large area ($\sim
37.5$ deg$^{2}$) to ensure statistically significant overlap in terms of
object numbers, yet is sufficiently small for all objects ($B_{\rm MGC}<20$
mag) to have been manually inspected (for all non-stellar objects) and
corrected, providing a robust and reliable survey. The MGC also probes to a
substantially deeper isophote ($\mu_{\rm B,lim} = 26$ mag arcsec$^{-2}$, Liske
et al. 2003) than either the original APM plate scans (upon which the 2dFGRS
input catalogue is based) or the SDSS drift scans (both with $\mu_{\rm B,lim}
\approx 24.5$ mag arcsec$^{-2}$, Maddox et al. 1990a, York et al. 2000). The
resulting higher signal-to-noise allows more reliable photometric measurement,
object classification and (de)-blending fixes. The deeper isophote also allows
a fully independent assessment of the completeness with regard to low surface
brightness galaxies (see Impey \& Bothun 1997). Likewise the higher resolution
and better mean seeing allows an assessment of the completeness for high
surface brightness galaxies (see for example Drinkwater et al. 1999).  These
latter concerns are aired in detail in Sprayberry et al. (1997; see also 
O'Neil \& Bothun 2000) who, following on from Disney (1976), argue for 
incompleteness
levels of as much as $50$ per cent in nearby galaxy catalogues such as the
APM. If this is indeed correct then a deeper survey such as the Millennium
Galaxy Catalogue should uncover a significant number of galaxies either missed
or with fluxes severely underestimated by the shallower surveys.

In this paper we describe the three independent imaging surveys (MGC, 
2dFGRS and SDSS-EDR) and the catalogue matching
process in Section 2 \& 3. We quantify the photometric accuracy as a function 
of magnitude and surface brightness in Section 4, including recent updates 
to the 2dFGRS including SuperCosmos data and the recently released SDSS 
First Data Release, and we quantify the reliability of the star-galaxy 
separation in Section 5. Finally we explore the crucial question 
of completeness across the apparent magnitude apparent surface brightness
plane ($M-\Sigma$) in Section 6. We summarise our findings in Section 7.

The Millennium Galaxy Catalogue is a publicly available dataset found at:
{\tt http://www.eso.org/$\sim$jliske/mgc/} or by request to 
jliske$\makeatletter{@}$eso.org (see Liske et al. 2003).

\section{Data}
Here we briefly introduce the three imaging catalogues which we 
wish to compare: The Millennium Galaxy Catalogue (MGC; Liske et al. 2003), 
the Two-degree Field Galaxy Redshift Survey (2dFGRS; Colless et al. 2001, 
2003)
and the Sloan Digital Sky Survey (Stoughton et al. 2002, Abazajian et al. 
2003). The MGC is adopted as the yardstick against which we shall quantify 
the photometric accuracy, completeness and contamination of the 2dFGRS 
and SDSS datasets. This is for a number of reasons:

\noindent
(1) The internal accuracy of the MGC photometry is shown to be $\pm 0.023$ 
mag, for stars and galaxies over the magnitude range $16 - 21$. This is 
superior
to the quoted accuracies of the 2dFGRS and SDSS-EDR datasets ($\pm 0.15$ mag,
Norberg et al. 2002b and $\pm 0.033$ mag, Stoughton et al. 2002, respectively).
\noindent
(2) The MGC is the deepest, in terms of sky noise, of the three surveys, 
extending to  $\mu_{\rm B_{\rm MGC},lim}=26.0$ mag arcsec$^{-2}$ (c.f. 
$\mu_{\rm b_{ J},lim}^{\rm 2dFGRS}=24.67$ mag arcsec$^{-2}$, Pimblett et al. 
2001 and $\mu_{\rm g^*,lim}^{\rm SDSS-EDR}=24.3$ mag arcsec$^{-2}$, 
Stoughton et al. 2002).

\noindent
(3) The MGC uses a fixed isophotal detection limit ensuring uniform survey 
completeness.

\noindent
(4) The MGC has the best median seeing of the three surveys with 
$\langle{\sc fwhm}_{\rm MGC}\rangle = 1.3''$ (cf. $\langle{\sc fwhm}_{\rm 2dFGRS} \rangle \sim 2.5''$ and 
$\langle{\sc fwhm}_{\rm SDSS-EDR}\rangle \sim 1.6''$).

\noindent
(5) All galaxies in the MGC to $B_{\rm MGC} = 20$ mag have been eyeballed and 
where necessary corrected for classification errors (overblending, underblending).

\noindent
(6) All CCDs have been carefully inspected and artefacts masked out 
(including satellite trails, hot pixels, bad columns and diffraction spikes).
Bright stars (on and off the image) and bright galaxies have also been masked
out, and asteroids and cosmic rays have been carefully identified,
see Liske et al. (2003). 
~

\begin{figure}
\vspace{-1.0cm}
\centerline{\psfig{file=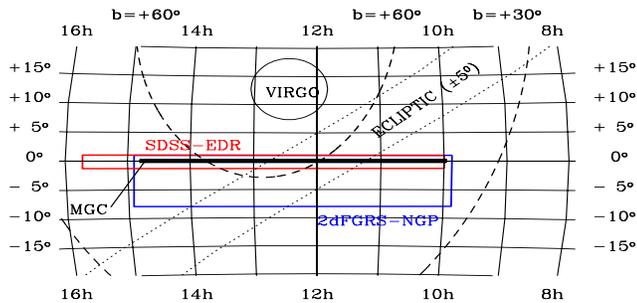,height=8.0cm,width=10.0cm}}
\vspace{-3.0cm}
\caption{A section of an Aitoff projection showing the overlap between
the three surveys covered in this paper. Also shown as dashed lines are
contours of constant Galactic Latitude, the location of the Ecliptic and
the Virgo cluster.}
\label{fig:aitoff}
\end{figure}

\noindent
Fig.~\ref{fig:aitoff} shows the region of overlap between the three surveys 
resulting in a
$\sim$ 30 deg$^{2}$ region in common comprising of $\sim 10,000$ MGC
galaxies in the magnitude range $16 < B_{\rm MGC} < 20.0$ mag (see Table 1).

\subsection{The MGC}
\label{sec:MGC}

The Millennium Galaxy Catalogue (Liske et al. 2003; henceforth MGC1) 
is a deep $\sim37.5$deg$^{2}$
B-band CCD imaging survey. It spans the 
equatorial strip from $9^h58^m28^s$ to 
$14^h46^m45^s$ with a declination range from $-0^{\circ}17'15''$ to 
$+0^{\circ}17'15''$ (J2000.0). 
The imaging was carried out using the 4-CCD mosaic Wide 
Field Camera on the Isaac Newton Telescope between March 1999 and April 2000 
and consists of 144 overlapping pointings.  The data were taken during dark
time with a median seeing {\sc fwhm}$=1.3''$, 
with pixel size $0.333''$, all objects are therefore fully sampled. Full 
details of the data 
collection, photometric and astrometric solutions along with the image 
detection, analysis and eyeball classifications are given in MGC1 and
summarised briefly below.

Objects were detected using Source Extractor ({\sc SExtractor}, Bertin \& 
Arnouts 1996) with a fixed isophotal detection threshold of 
$\mu_{\rm B_{\rm MGC},lim}=26$ mag arcsec$^{-2}$. The final MGC magnitudes 
are dust corrected (Schlegel, Finkbeiner \& Davis 1998) SExtractor ``best'' 
magnitudes, which are derived from an elliptical aperture of 2.5 Kron Radii 
(Kron 1980), unless the nearby neighbour 
flag is set in which case the Gaussian corrected isophotal magnitude is used.
From the overlap regions between adjacent pointings we have determined
that the internal astrometric and photometric error distributions are well 
described by Gaussians of {\sc fwhm} $\pm 0.08''$ and $\pm 0.023$ mag 
respectively (see Fig. 5 \& 7 of MGC1). The calibration solution from Landolt 
standards indicates that the absolute zero point is accurate to 
$\pm 0.005$ mag.

The MGC was separated into two magnitude ranges forming the
{\sc MGC-BRIGHT} ($16.0 < B_{\rm MGC} < 20.0$ mag) and {\sc MGC-FAINT} 
($20.0 \leq B_{\rm MGC} < 24.0$ mag) catalogues. For the purposes of this paper we
now focus on {\sc MGC-BRIGHT}. We note that the {\sc stellaricity}
distribution is extremely bimodal indicating reliable star-galaxy separation,
(see Fig 9 of MGC1) even so all objects with
{\sc stellaricity} $< 0.98$ were visually inspected, classified and where 
necessary repaired manually for erroneous deblending, erroneous 
background estimation or contamination from nearby objects. A flag was 
assigned for each galaxy indicating whether its photometry was considered
``good'', ``compromised'' or ``corrupted''.

Bright stars, diffraction spikes, bad columns, hot pixels, 
satellite trails, bad charge transfer regions and CCD edges were masked. 
Objects within a 50 pixel threshold of a masked pixel were removed from 
the catalogue to produce a final pristine fully eyeballed catalogue of 
9795 galaxies (9657 ``good'', 137 ``compromised'' and 0 ``corrupted'') 
within a reduced survey area of 30.90 deg$^{2}$ over the magnitude range 
$16 < B_{\rm MGC} < 20$ mag.

Half-light radii were measured for all galaxies within {\sc MGC-BRIGHT} 
and are equal to the semi-major axis of the 
ellipse that contains half of the flux of the galaxy. The effective surface 
brightness is then derived assuming a circular aperture
(i.e., $\mu_{\rm eff} = B_{\rm MGC}+2.5\log_{10}[2 \pi r_{\rm hlr}^{2}]$). 
If the galaxy is an 
inclined optically thin disk galaxy, this will correct the effective surface 
brightness to the face on values (see Cross \& Driver 2002 for a more detailed
discussion of the implications of this).

\subsection{The 2dFGRS}
\label{sec:2dF}

The 2dFGRS contains both photometric and 
spectroscopic data for 229,118 galaxies selected from the Automated Plate 
Measuring-machine galaxy catalogue (APM; Maddox et al. 1990a,b). The 2dFGRS 
target catalogue covers $2152\deg^2$ to a limiting magnitude of 
$b_{\rm j,old} = 19.45$ mag, where 
$b_{\rm j,old}$ is the photometry of the galaxies at the beginning of the 
2dFGRS campaign, before photometry updates in 2001 and 2003.  

We will briefly describe the calibration process here, but the full 
calibration and recalibration up to and including the 2001 recalibration is
described in detail in Maddox et al. (1990a) and Norberg et al. (2002b). 

The APM images come from photographic plates collected on the UK Schmidt 
Telescope 20 to 30 years ago and digitised by the APM team. The APM 
magnitudes were measured with an approximate surface brightness 
limit of $\mu_{\rm b_j,lim}\sim24.67$ mag arcsec$^{-2}$ (see Cross 
et al. 2001; Pimblett et al. 2001). The original isophotal magnitudes were 
adjusted assuming a Gaussian profile to produce pseudo-total 
magnitudes (see Maddox et al. 1990b for details). For
a subsample total CCD magnitudes were obtained and converted to the 
$b_{\rm j}$ band using $b_{\rm j}=B-0.28(B-V)$ (Blair \& Gilmore 1982). 
A calibration curve was determined by minimising the residuals between
the APM $b_{\rm j}$ and the CCD $b_{\rm j}$. 

When the 2dFGRS target catalogue was determined, in 1994, more CCD data 
were available (see Norberg et al. 2002b). New offsets between the original 
field corrected total 
magnitudes and the final magnitudes were obtained, assuming a fixed scale 
of 1, to select the sample. While any scale error will produce errors in
the bright magnitudes, it will not affect the selection of targets at 
$b_{\rm j}=19.45$. Additional UKST plates outside the APM Galaxy Survey were 
reduced using the standard APM galaxy survey procedures to improve the 
efficiency of the 2dFGRS observing strategy. These additional fields contained
data in the region $9^h<{\rm R.A.}<15^h$ and  
$-7^{\circ}.5<{\rm DEC}<3^{\circ}.5$ (J2000.0).
This additional data were calibrated separately using CCD data from 
Raychaudhury et al. (1994) and contains the data used in this paper. The 
magnitudes were then dust corrected using the dust maps supplied by
David Schlegel, similar to the maps in Schlegel, Finkbeiner \& Davis (1998).

\onecolumn

\begin{figure}
\centerline{\psfig{file=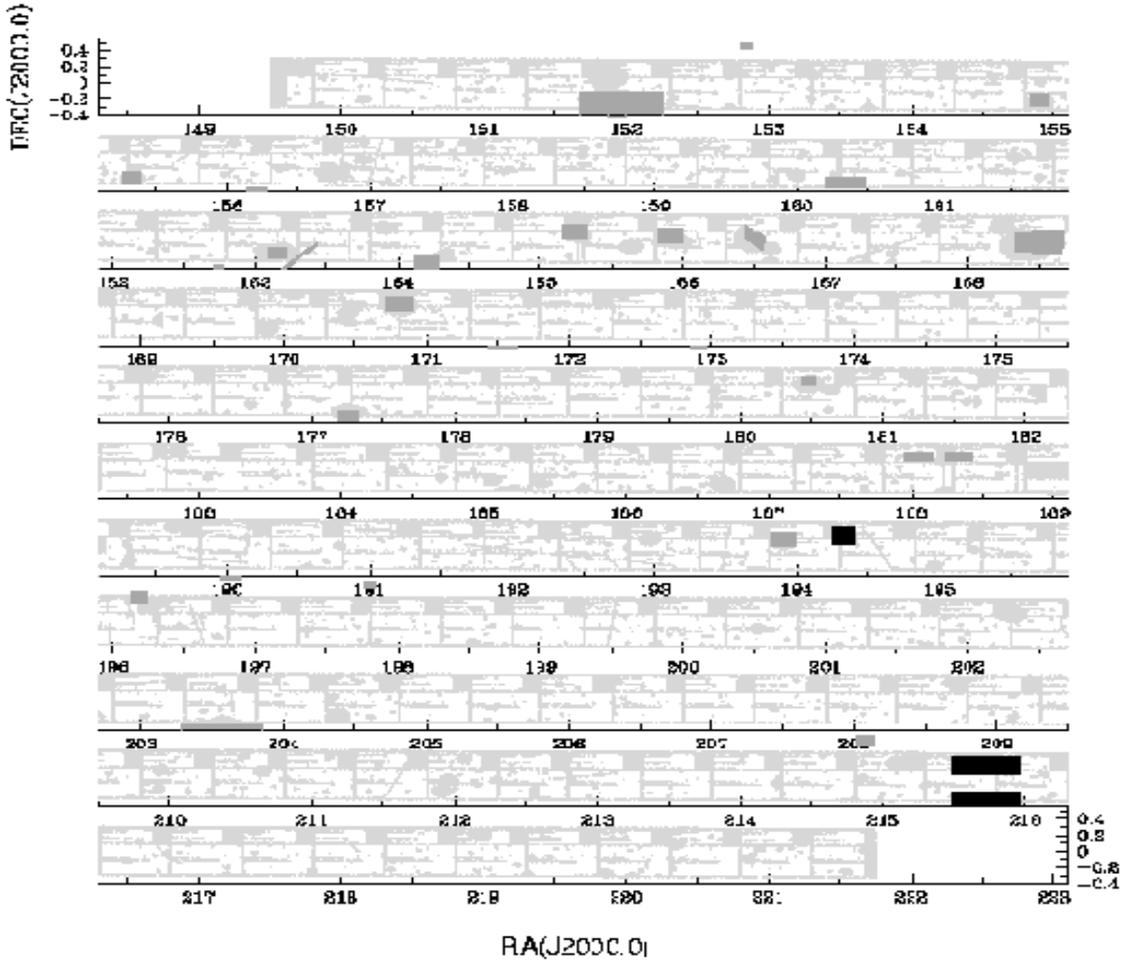,height=200mm,width=\textwidth}}
\caption{The common region for the three surveys. The light grey, 
dark grey and 
black regions represent exclusion regions from the MGC, 2dFGRS and SDSS-EDR
respectively. The remaining white area covers 29.74 deg$^{2}$ and represents
the region in common between the three imaging surveys.}
\label{fig:masks}
\end{figure} 

\twocolumn

In 2001, the subset of the APM representing the 2dFGRS input catalogue was
recalibrated further using European Imaging Survey Data (Arnouts et al. 2001)
to provide an absolute revised zero point for plate number UKST 411.  The
($b_{\rm j} - J)$ colour vs $b_{\rm j}$ relation was then derived for UKST 411
using $J$-band data from the 2-Micron All Sky Survey (2MASS, Jarrett et
al. 2000). However, this revision does not affect the target selection in the
2dFGRS, and it is not relevant to this paper.

In April 2003, the photometry was recalibrated once again by comparing
magnitudes calculated from APM scans with magnitudes calculated from
SuperCosmos (SCOS, Hambly et al. 2001) scans (see Peacock et al. 2003, 
Colless et al. 2003 for details).
The SCOS data were calibrated from external CCD sources (mainly SDSS 
EDR with updated zero points), but with the mean 2MASS ($b_{\rm j}-J$) on each
plate forced to be the same. The UKST $b_{\rm j}$ and $r_{\rm f}$ plates were
calibrated separately, but then a final iteration was performed to keep the
distribution of $b_{\rm j}-r_{\rm f}$ colours uniform. Finally the original
APM data were regressed to fit the SuperCosmos data for each plate.  The
absolute precision of the photometry is limited by the uniformity of the 2MASS
photometry which is claimed to be good to 0.03 mag over the whole sky.

The SCOS data includes both $b_{\rm j}$and $r_{\rm f}$ magnitudes from the 
same plates as
the APM data, but with independent scanning and calibration. These are 
included in the main 2dFGRS data base, see Colless et al. (2003). In Section 4
we test the photometry of SCOS as well as 2dFGRS against the MGC.

The final dust-corrected 2dFGRS $b_{\rm j}$ magnitudes will be referred to as 
$b_{\rm 2dF}$ throughout. Since the 2dFGRS selection limit was 
$b_{\rm 2dF,old} = 19.45$ 
mag, the redshift survey does not have a fixed limiting magnitude. The SCOS 
$b_{\rm j}$ and $r_{\rm f}$ magnitudes will be referred to as $b_{\rm SCOS}$ 
and $r_{\rm SCOS}$.

Star-galaxy separation was implemented as described in Maddox et
al. (1990a). They estimate that the star classification is reliable with 
$\sim 5$ per cent stellar contamination to a limit of 
$b_{\rm 2dF,old} \approx 20.4$ mag.

\subsection{The SDSS}
\label{sec:SDSS}

We use data from the Sloan Digital Sky Survey Early Data Release (SDSS-EDR)
and Sloan Digital Sky Survey First Data Release (SDSS-DR1).
The SDSS-EDR (Stoughton et al. 2002)
consists of 8 drift scan stripes covering 3 regions obtained via a dedicated 
2.5m telescope at the Apache Point Observatory. The 2001 EDR-region covers a 
total of 462 deg$^{2}$ providing photometry in $u^*$, $g^*$, $r^*$, 
$i^*$ and $z^*$ for $\sim 14$ million objects to approximate point source 
detection limits of $22.0, 22.2, 22.2, 21.3,$ and $20.5$ mag respectively.
The SDSS-DR1 (Abazajian et al. 2003) covers 2099 deg$^{2}$, in $u$, $g$, $r$, 
$i$ and $z$ including the SDSS-EDR, with improvements to the data extraction. 
While these improvements include deblending, astrometry
and spectroscopy, the main improvements are in the photometry. Since most of
the deblending problems are for $r<15$ galaxies this will not significantly
affect our completeness. Therefore we have stuck to the SDSS-EDR in the 
completeness and contamination sections. 

The effective integration time of SDSS is 54 seconds yielding an approximate 
isophotal 
detection limit of $\mu_{\rm g^*,lim}=24.3$ mag arcsec$^{-2}$ and 
$\mu_{\rm r^*,lim}=24.1$ mag arcsec$^{-2}$. The data overlapping the MGC and 
2dFGRS (stripes 752 \& 756) were taken through variable conditions with 
seeing ranging from $1.0''$ to $3.0''$ (see Figure 8, Stoughton et al. 2002). 
Photometric 
calibration is made with the use of a nearby telescope to measure nightly 
extinction values and ``observation transfer fields'' which lie within the 
SDSS survey areas.

Image detection, analysis and classification was undertaken using in-house 
automated software producing a final set of 120 parameters or flags 
per object. Full details of the data reduction pipeline are given in Stoughton
et al., (2002) and references therein. Preliminary galaxy number-counts and 
discussion of the 
completeness and contamination at magnitudes brighter than $g^*=16$ mag are 
given in Yasuda et al. (2001). 

The final SDSS database defines a number of magnitude measurements and we 
shall adopt the reddening corrected Petrosian magnitudes (see 
Fukugita et al. 1996) as closest to 
total --- shown to have no surface brightness dependency for 
a well defined profile shape. The final quoted SDSS-EDR photometric accuracy 
is $\pm 0.033$ mag and the pixel size is $0.396''$. 

In the SDSS literature, there are many methods of star-galaxy classification. 
We have taken the classifications used in the SDSS-EDR database, which are 
calculated as prescribed in Stoughton et al. (2002). They separate stars from 
galaxies using the difference between the PSF and model magnitude in $r^*$. 
Galaxy target selection requires a difference greater than 0.3 mag. 

\subsection{Additional redshift data}

The redshift data from 2dFGRS and SDSS-EDR have also been supplemented with
6065 additional redshifts taken by the authors using the 2dF instrument. These
are the first part of a dataset designed to provide a complete sample of
galaxies with $B_{\rm MGC}<20.0$ mag, selected from the MGC and a complete
sample of stars with $B_{\rm MGC}<20.0$ mag for a section of the survey. We
have also added in 4007 redshifts from the NASA Extragalactic Database (NED),
736 redshifts from the 2dF QSO Redshift Survey (2QZ), 55 redshifts from Paul
Francis' Quasar Survey (Francis, Nelson \& Cutri 2003) and 11 Low Surface
Brightness Galaxy redshifts (Impey et al. 1996).  There are many galaxies 
for which we have 
multiple redshifts and we have a high overall completeness. Out
of the 9795 ($16<B_{\rm MGC}<20$ mag) MGC objects classified as galaxies, 8837
have redshifts, $90.2$ per cent. This proportion rises to $96.0$ per cent for 
$16<B_{\rm MGC}<19.5$ mag galaxies and $98.8$ per cent for 
$16<B_{\rm MGC}<19$ mag galaxies. There are
also 2907 MGC stellar-like objects ($16<B_{\rm MGC}<20$ mag) with measured
velocities.

\subsection{Filter conversions}
\label{sec:filcon}

We elect to work in the $B_{\rm MGC}$ band for which the following filter 
conversions have been derived (based upon Fukugita et al. 1996; Norberg et 
al. 2002b, Smith et al. 2002 and MGC1). The full details of this analysis are
found in the Appendix. The four colour equations are for 2dFGRS, SCOS, 
SDSS-EDR and SDSS-DR1 respectively. 

\begin{equation}
B_{\rm MGC} = b_{\rm 2dF} + 0.121(g^*-r^*) - 0.012 
\end{equation}

\begin{equation}
B_{\rm MGC} = b_{\rm SCOS} + 0.108(b_{\rm SCOS}-r_{\rm SCOS}) - 0.044 
\end{equation}

\begin{equation}
B_{\rm MGC} = g^* + 0.251(g^*-r^*) + 0.178 \label{eq:colSDMGC}
\end{equation}

\begin{equation}
B_{\rm MGC} = g + 0.251(g-r) + 0.178 
\end{equation}

We use the colours from each dataset
where possible. For the 2dFGRS we use the SDSS-EDR colours, since much of
the 2dFGRS calibration was done using the SDSS-EDR. We also tried using
the SCOS $r_{\rm f}$ data for the 2dFGRS colour equation. This gave similar 
photometric results when compared to other surveys, but with more scatter.

\subsection{Masking and areal coverage}
All three surveys contain unobserved regions due to a variety of issues
most notably bright stars (2dFGRS), failed scans (SDSS-EDR) and CCD 
cracks/boundaries (MGC). As the MGC is wholly contained within the 2dFGRS
and SDSS-EDR regions we trim all three catalogues to an approximate
44 deg$^{2}$ common range defined by the MGC: 
$9^h 58^m 00^s < \alpha_{\rm J2000} < 14^h 47^m 00^s$ and 
$-00^{\circ} 18' 00'' < \delta_{\rm J2000} < 00^{\circ} 18' 36''$. Within this 
rectangle
the MGC covers 37.5 deg$^{2}$ of which 30.9 deg$^{2}$ is considered
high quality. The SDSS-EDR contains three holes within this region and the
2dFGRS contains a number of star ``drill''-holes. Taking all exclusion
regions into account we are left with a final high quality fully covered
common area of 29.74 deg$^{2}$. The number of objects contained within this
common region for each of the three surveys is shown in Table 1.
Fig.~\ref{fig:masks} shows the common region with the individual masks 
overlaid for the MGC (light grey), 2dFGRS (dark grey) and SDSS-EDR (black) 
surveys respectively.

\begin{table}
\caption{Summary of objects and depth within the common area for the three
surveys}
\begin{tabular}{lccc}
Survey & Range (mag) & No. Gals. & No. Stars \\ \hline \hline
MGC-BRIGHT & $16.0 < B_{\rm MGC} < 20.0 $  &  9,795 & 36,260 \\
SDSS-EDR   & $16.0 < B_{\rm MGC} < 20.0 $  & 10,213 & 34,779 \\
2dFGRS     & $16.0 < B_{\rm MGC} < 19.5 $ &  5,215 & ---    \\
MGC-BRIGHT & $16.0 < B_{\rm MGC} < 19.5 $ &  5,792 & 28,271 \\
SDSS-EDR   & $16.0 < B_{\rm MGC} < 19.5 $ &  6,063 & 27,026\\ \hline
\end{tabular}
\end{table}

\section{Catalogue Matching}

\subsection{Matching MGC to 2dFGRS}

\begin{figure}
{\psfig{file=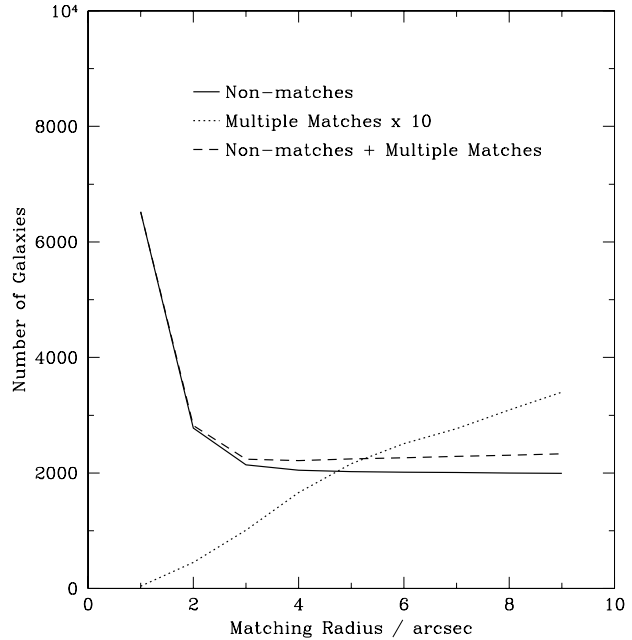,width=90mm,height=90mm}}
\caption[Optimal matching radius for galaxies]{The solid line shows the 
number of 2dFGRS galaxies {\it not} matched to MGC galaxies. The dotted line
shows the number of multiple matches multiplied by 10, and the dashed line
shows the sum of the non-matches and multiple matches.}
\label{fig:matchrad}
\end{figure}

The catalogue of 
2dFGRS objects described in Section~\ref{sec:2dF} was matched to the {\sc 
MGC-BRIGHT} catalogue ($B_{\rm MGC}<20.0$ mag) by finding the nearest match 
within a radius of $5\arcsec$. Various radii were tested, see 
Fig~\ref{fig:matchrad}. While the minimum sum of the non-matches and multiple 
matches is $4\arcsec$, most close-in multiple matches occur when a 2dFGRS
object is composed of two or more MGC matches (see Cross 2002 for more
details) rather than a nearby unassociated object being wrongly matched. The
gradient of the number of multiple objects reaches a maximum at $5\arcsec$,
indicating that most multiples are due to poor resolution within this radius.
Thus $5\arcsec$ gives the optimal radius to maximise the number of real 
matches.

Each 2dFGRS object was also checked for multiple matches within an ellipse 
defined by its isophotal area, eccentricity and orientation. If an MGC object 
contributes to the flux of the 2dFGRS object, its centre should lie within 
the area of the 2dFGRS object. To find multiple and faint matches, objects in 
{\sc MGC-FAINT} with $B_{\rm MGC}<21.0$ mag were also matched. The edge, $s$ 
of the 2dFGRS object is defined below.

\begin{equation}
s=\left[\left(\frac{\cos(\theta)}{a}\right)^2+\left(\frac{\sin(\theta)}{b}\right)^2\right]^{-\frac{1}{2}}
\end{equation}

\noindent where $a$ is the length of the semi-major axis, $b$ is the length 
of the semi-minor axis and $\theta$ is the bearing 
from the 2dFGRS object to the MGC object and the orientation of the 2dFGRS 
object on the sky. $a$ and $b$ are defined from the area ($A$) and the 
eccentricity ($e$) below.

\begin{equation}
a=\sqrt{\left(\frac{A}{\pi}\right)}(1-e^2)^{-0.25} 
\end{equation}

\begin{equation}
b=\sqrt{\left(\frac{A}{\pi}\right)}(1-e^2)^{0.25} 
\end{equation}

\begin{figure}
{\psfig{file=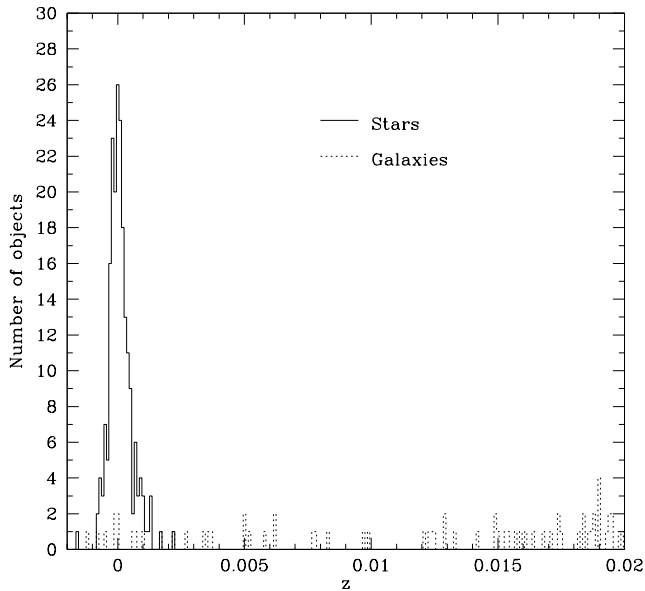,width=90mm,height=90mm}}
\caption[$N(z)$ at low redshift, showing stars and galaxies]{Histograms of 
the redshift distributions of stars (solid line) and 
galaxies (dotted line) of 2dFGRS objects at low redshift. Stars have a narrow 
distribution with width $\sim1.0\times10^{-3}$ centred on $z=0$.}
\label{fig:lowz}
\end{figure}

If the MGC object lies at $r\leq\,s$ then it is a component of the 2dFGRS object. The main component is deemed to be the brightest, unless the 
redshift is incompatible with the MGC star-galaxy classification. Using both 
methods allows for some error in the position and picks up almost all the 
matches first time. 
A few (5) matches were missed by both methods, because they were slightly
too wide or slightly too faint. These were put in later by hand.
Fig.~\ref{fig:lowz} shows single component MGC-2dFGRS matches at $z<0.02$. 
The solid histogram shows the objects that are classified as stars in the MGC 
and the dotted histogram shows the objects that are classified as galaxies in 
the MGC. It is clear that the stellar population has a distribution with 
$z\leq\,2.0\times10^{-3}$, at which redshift there are very few galaxies. 
For multiple matches the MGC comparison magnitude is taken as the sum of all
components lying within the 2dFGRS objects isophotal area. All failed matches 
were checked by eye. Many were objects lying close to the exclusion boundaries.
The 2dFGRS contains 5,346 objects within the common region ($B_{\rm MGC}>16.0$ 
mag) of which matches are found for 5,285 and 61 have no matches. Of these 
mismatches 8 are due to the MGC objects lying across an exclusion boundary 
(close to bright stars) and 53 are genuine mismatches. 4 of these are due
to overdeblending of very bright galaxies by the APM process and 49 have no 
obvious counterparts on the MGC data and must represent plate 
artefacts, asteroids, satellite trails, diffraction spikes or other such 
objects. These are described in Section~\ref{sec:art}.

The 2dFGRS-MGC catalogue was then inverted so that the MGC was the reference
catalogue. If the MGC object had $B_{\rm MGC}\ge20$ mag it was removed. All 
additional components from the matching done above, with $B_{\rm MGC}<20$ mag, 
were also added in. Finally each MGC galaxy was checked for multiple 2dFGRS 
objects within an ellipse defined by the MGC ellipticity, isophotal area and 
position 
angle. There are 46,364 MGC objects in the common region, of which 9,795 are 
classified as galaxies, 36,260 are classified as stars and 309 are classified 
as asteroids, cosmic rays, noise detections or obsolete (see Liske et al.
2003). Of the 9,795 galaxies, 4,646 have a single match to 2dFGRS galaxies, 405
have two or more MGC objects (of which the brightest is a galaxy) matched to 
one 2dFGRS objects, 2 have two or more 2dFGRS objects matched to one MGC 
galaxy, and 4,742 have no match, mainly because the MGC limiting magnitude is 
fainter than the 2dFGRS limits.

\subsection{Matching the SDSS-EDR to the MGC} 

A similar strategy to the above was employed for the matching of the MGC and 
SDSS-EDR catalogues. The only exception was in the handling of multiple-matches
where the MGC data could not be guaranteed to have superior resolution in all 
cases. Hence for multiple-matches we also employ a nearest neighbour routine.
This produces a match if and only if galaxy A is the nearest object to galaxy
B and galaxy B is the nearest object to galaxy A. It also identifies secondary
components as objects where a second galaxy C has A as the nearest match. For 
each galaxy we find the neighbours using both methods. The components of a 
particular galaxy are those selected by both routines. There are 44,992 SDSS 
objects ($16<g^*+0.251(g^*-r^*)+0.178<20$ mag) in the common region, 
of which 10,213 are classified as galaxies and 34,779 are classified as stars. 
Of the 10,213 galaxies, 9039 have clear matches to MGC galaxies, 18 have 
multiple matches to MGC objects (of which the brightest is a galaxy), 260 
have matches to star-like MGC objects, and 858 have no match. Of the 34,779
stars, 34,213 have matches to single MGC stars, 35 have multiple matches, with
the brightest matching to a star, 9 are matched to non-stellar objects, 
and 521 have no match. After comparison to {\sc MGC-FAINT}, the non-matches
reduced to 329 galaxies and 335 stars. These are discussed in Section~\ref{sec:art}.

\section{Photometric Comparison}

All the following numbers are selected with $16<B_{\rm MGC}<20$ mag to avoid 
problems with saturation at the bright end. After matching to the MGC we find 
unambiguous single-single object matches in the common region for 4,418 
2dFGRS objects and 44,690 SDSS-EDR objects and a further 589 
($11.7$ per cent) and 893 ($2.0$ per cent) ambiguous or multiple 
matches respectively. The ambiguous matches are MGC galaxies matched to 
2dFGRS/SDSS-EDR stars or vice-versa. For the purposes of photometric 
comparisons we now consider only the unambiguous single-single object 
matches. However first it is worth 
considering the various magnitudes used in this section. The MGC
adopts Kron magnitudes (Kron 1980) defined by an elliptical aperture of 
major axis 2.5 Kron radii and ellipticity as defined by the initial
SExtractor parameters. The 2dFGRS uses isophotally corrected magnitudes
with subsequent corrections for zero-point offsets and scale-errors
(see Norberg et al. 2002b).
The SDSS-EDR uses Petrosian magnitudes (see Blanton et al. 2001). 
All magnitude systems have their virtues and 
failings and we defer a prolonged discussion of this by simply
choosing to compare the final quoted magnitudes for each survey as seen 
by the user. As a reminder we note that Petrosian magnitudes are known to 
underestimate the total magnitudes for Gaussian/exponential/de Vaucouleurs 
profiles by 0.07, 0.01 and 0.22 mag respectively whereas Kron magnitudes 
are known to underestimate the same profiles by 0.01, 0.04 and 0.10 mag 
respectively (see Cross 2002). The 2dFGRS isophotally-corrected magnitudes 
are deemed total and no quantifiable error for profile shapes is known.

Hence our comparison will naturally incorporate discrepancies in photometry, 
methodology and spectral shape assumptions in the colour conversions 
resulting in a ``real-life'' assessment of the error budget. While the 
dust correction is part of the ``real-life'' assessment, it varies as a 
function of position, and so must be dealt with separately. Therefore we do 
the photometry on magnitudes uncorrected for extinction throughout. 

The extra dust correction terms are important since all the magnitudes have 
been dust 
corrected independently albeit 
based upon the same dust-maps (Schlegel, Finkbeiner \& Davis 1998). It was
found that the different values of ${\rm A/E(B-V)}$ were slightly 
inconsistent and contributed the following additional offsets to the data:

\begin{eqnarray}
&\Delta\,B{\rm ^{DC}(MGC-2dFGRS)}=-0.002 \\
&\Delta\,B{\rm ^{DC}(MGC-SDSS)}=-0.007 \\
&\Delta\,B{\rm ^{DC}(2dFGRS-SDSS)}=-0.005
\end{eqnarray}

 The corrections 
for 2dFGRS
and SCOS are the same and the corrections for the SDSS-EDR and SDSS-DR1 are 
the same. The dust corrections do not appear to increase the variance. This 
additional offset is directly proportional to the mean dust correction over 
the survey strip, ${\rm \overline{E(B-V)}}=0.033\pm0.010$.  The
additional offsets can be calculated in other parts of the sky using the 
following equation:

\begin{equation}
\Delta\,B^{\rm DC}=\Delta\,{\rm [A/E(B-V)] \overline{E(B-V)}}
\end{equation}

\noindent where $\Delta\,{\rm [A/E(B-V)]}=0.06,0.21,0.15$ for (MGC-2dFGRS), 
(MGC-SDSS) and (2dFGRS-SDSS) respectively.
The SCOS and SDSS-DR1 photometry is added to the 2dFGRS and SDSS-EDR matches
respectively. The full summary of all the cross-checks is given in 
Tables~\ref{tab:phot}, ~\ref{tab:phstars} and ~\ref{tab:phtsb} which list
galaxy scale errors, stellar photometry and errors with surface brightness  
respectively.

\onecolumn

\begin{table*}
\caption{Summary of the relative photometry of galaxies. This table lists 
the offset, average standard deviation per galaxy and the parameters $a$ and $b$ from Eqn~\ref{eq:scl}. We list each set of numbers for the full range of magnitudes, for the 3-way cross check ($16<B_{\rm MGC}<19$ and for the faint 
sample $19<B_{\rm MGC}<20$.}
{\footnotesize
\begin{tabular}{lrrrrrrrrrrrrr}\label{tab:phot}
Data & \multicolumn{5}{|c|}{Full Range of Data} & \multicolumn{4}{|c|}{Best Sample $16<B<19$} & \multicolumn{4}{|c|}{Faint Sample $19<B<20$} \\ 
    & mean & $\sigma$ & a & b & $\chi^2_{\nu}$ & mean & a & b & $\chi^2_{\nu}$ & mean & a & b  & $\chi^2_{\nu}$ \\
\hline \hline
MGC-2dF  &  0.035 & 0.142 &  0.114 &  0.057 & 1.28 & 0.012 &  0.099 &  0.051 & 1.45 &  $---$ &  $---$ &  $---$ & $--$ \\
MGC-SCOS & 0.032 & 0.108 &  0.096 &  0.045 & 3.28 & 0.017 &  0.093 &  0.044 & 4.39 &  $---$ &  $---$ &  $---$ & $--$ \\
MGC-EDR  & 0.032 & 0.094 & 0.045 &  0.014 & 9.72 & 0.026 &  0.073 &  0.026 & 3.33 & 0.036 & 0.036 & $-$0.002 & 1.03 \\
MGC-DR1  & 0.039 & 0.086 &  0.051 &  0.013 & 14.0 & 0.034 &  0.082 &  0.027 & 5.06 & 0.042 & 0.040 & $-$0.006 & 1.01 \\
2dF-DR1  & 0.004 & 0.151 & $-$0.044 & $-$0.037 & 1.14 &  0.019 & $-$0.029 & $-$0.028 & 0.60 &  $---$ &  $---$ &  $---$ & $--$ \\
SCOS-DR1 &  0.005 & 0.104 & $-$0.024 & $-$0.021 & 0.87 &  0.013 & $-$0.017 & $-$0.018 & 0.69 &  $---$ &  $---$ &  $---$ & $--$ \\
DR1-EDR  & $-$0.007 & 0.037 & $-$0.008 & $-$0.000 & 1.98 & $-$0.008 & $-$0.011 & $-$0.002 & 1.41 & $-$0.007 & $-$0.007 &  0.000 & 1.53 \\
SCOS-EDR & $-$0.002 & 0.107 & $-$0.035 & $-$0.023 & 0.75 &  0.006 & $-$0.031 & $-$0.021 & 0.79 &  $---$ &  $---$ &  $---$ & $--$ \\
2dF-EDR  & $-$0.003 & 0.154 & $-$0.052 & $-$0.037 & 1.04 &  0.011 & $-$0.038 & $-$0.029 & 0.59 &  $---$ &  $---$ &  $---$ & $--$ \\
SCOS-2dF &  0.001 & 0.086 &  0.018 &  0.012 & 1.71 & $-$0.006 &  0.009 &  0.009 & 0.93 &  $---$ &  $---$ &  $---$ & $--$ \\ \hline
\end{tabular}
}
\end{table*}

\begin{table*}
\caption{Summary of the relative photometry between stellar objects. The 
columns listed give the mean, standard deviation per star, scale error (see 
Eqn~\ref{eq:scl}) and the aperture corrected mean and standard deviation 
corrected for field-to-field variation.}
\begin{tabular}{l|rrrrr|rr}\label{tab:phstars}
Data & mean & $\sigma$ & a & b & $\chi^2_{\nu}$ & Best mean$^1$ & Best $\sigma^2$  \\
\hline \hline
MGC-EDR  & $-$0.031 & 0.057 & $-$0.038 & $-$0.004 & 96.6 & 0.035 &  0.046 \\
MGC-DR1  & $-$0.022 & 0.057 & $-$0.028 & $-$0.004 & 90.3 & 0.044 &  0.046 \\
\hline
\end{tabular}

\break 
\noindent $^1$ The best mean has aperture corrections removed. There is a 
0.066 mag correction for Petrosian to Kron magnitudes.
\break
\noindent $^2$ The best $\sigma$ has field-to-field variations removed.
\end{table*}

\begin{table*}
\caption{Summary of photometric differences as a function of surface 
brightness. This table lists the offset, average standard deviation per 
galaxy, the best linear-fit parameters $a$ and $b$ from Eqn~\ref{eq:sblin}
and the best quadratic fit parameters $\alpha$, $\beta$ and $\gamma$ from 
Eqn~\ref{eq:sbqud}.}
\begin{tabular}{l|rr|rrr|rrrr}\label{tab:phtsb}
Data & mean & $\sigma$ & a & b & $\chi^2_{\nu}$ & $\alpha$ & $\beta$ & $\gamma$ & $\chi^2_{\nu}$ \\
\hline \hline
MGC-2dF  & 0.035 & 0.142 &  0.053 &  0.096 & 35.1 &  0.073 &  0.081 & $-$0.041 & 1.44 \\
MGC-SCOS & 0.032 & 0.108 & 0.040 &  0.026 & 37.5 &  0.057 &  0.012 & $-$0.036 & 1.65 \\
MGC-EDR  & 0.032 & 0.094 & 0.031 & $-$0.016 & 22.5 & 0.039 & $-$0.025 & $-$0.015 & 1.29 \\
MGC-DR1  & 0.039 & 0.086 & 0.039 & $-$0.018 & 28.0 & 0.039 & $-$0.026 & $-$0.016 & 1.09 \\
2dF-DR1  & 0.004 & 0.151 & $-$0.018 & $-$0.126 & 15.9 & $-$0.031 & $-$0.113 &  0.026 & 2.15 \\
SCOS-DR1 & 0.005 & 0.103 & $-$0.007 & $-$0.056 & 17.0 & $-$0.016 & $-$0.043 &  0.022 & 2.22 \\
DR1-EDR  & $-$0.007 & 0.037 & $-$0.008 &  0.0003& 2.11 & $-$0.008 &  0.002 &  0.001 & 1.06 \\
SCOS-EDR & $-$0.002 & 0.107 & $-$0.015 & $-$0.057 & 17.3 & $-$0.024 & $-$0.043 &  0.022 & 3.94 \\
2dF-EDR  & $-$0.003 & 0.154 & $-$0.024 & $-$0.125 & 16.1 & $-$0.037 & $-$0.113 &  0.026 & 2.69 \\
SCOS-2dF &  0.001 & 0.086 &  0.012 &  0.070 & 8.10 &  0.015 &  0.068 & $-$0.006 & 5.83 \\ \hline

\end{tabular}
\end{table*}

\twocolumn

\subsection{Magnitude offset and scale-errors}

\begin{figure}
\centerline{\psfig{file=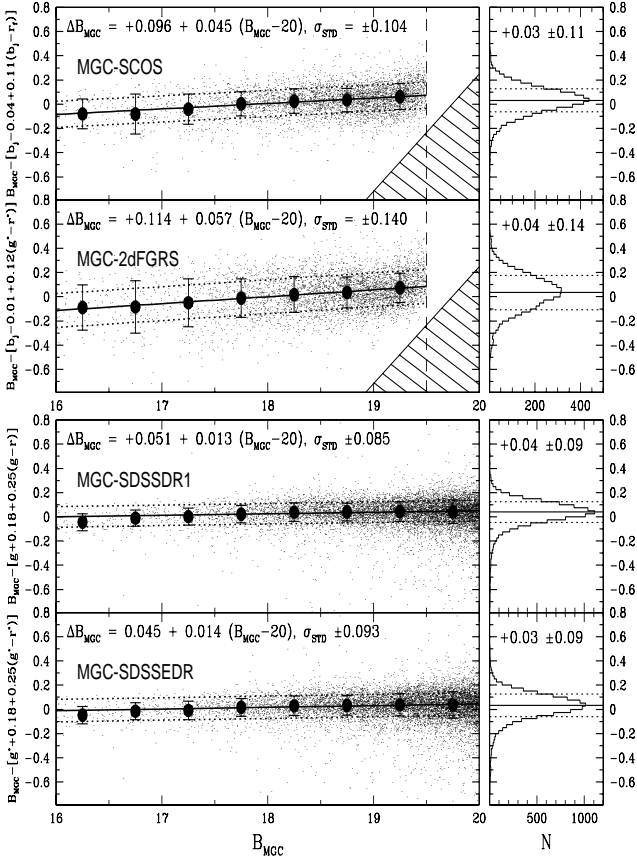,height=120.0mm,width=90.0mm}}
\caption{A comparison of photometry between MGC and SCOS (top panel), 2dFGRS 
(second panel), SDSS-DR1 (third panel) and SDSS-EDR (bottom panel). 
The left panels show the magnitude dependence
along with the robust estimate of the best-fit-line via a chi-squared 
fit to the $3-\sigma$ clipped standard deviation in each bin. The standard
deviation quoted with the scale-error is the standard deviation after
subtracting this scale-error. The right-hand panels show the histogram of the
photometric differences with the $3-\sigma$ clipped mean and 
standard-deviations marked. See Table~\ref{tab:phot} for more details.}
\label{fig:phot_sing}
\end{figure} 

\noindent
Fig.~\ref{fig:phot_sing} shows the photometric comparison between good quality
single-single matches within the common area for $B_{\rm MGC}-B_{\rm SCOS}$
(top), $B_{\rm MGC}-B_{\rm 2dF}$ (second from top), $B_{\rm MGC}-B_{\rm 
SDSS-DR1}$ (third from top) and $B_{\rm MGC}-B_{\rm SDSS-EDR}$ (bottom).  
Note that the 2dFGRS, SCOS, SDSS-EDR and SDSS-DR1 magnitudes were transformed 
according to the colour equations in Section~\ref{sec:filcon} using the 
appropriate SDSS colour or SCOS colour for each individual galaxy. The 
photometric differences 
between the surveys are summarised in Table~\ref{tab:phot}. The mean error in 
the photometry is $\sim0.035$ mag for the comparison of MGC to the other 
surveys with less than $0.01$ mag difference between these other surveys.
Since 2dFGRS and SCOS have been calibrated to the SDSS-EDR, this latter 
result is not surprising. $B_{\rm MGC}-B_{\rm 2dF}=0.035\pm0.005$\footnote{
While the random errors between the MGC and 2dFGRS account for an error of 
0.002 mag only, the colour equations in 
Fukugita et al. (1996), Smith et al. (2002) and Blair \& Gilmore (1982) are 
only quoted to 2 decimal places, $\pm0.005$ mag.} mag with a standard 
deviation per galaxy of 0.142 mag, $B_{\rm MGC}-B_{\rm SCOS}=0.032\pm0.005$ 
mag with a standard deviation per galaxy of 0.108 mag, 
$B_{\rm MGC}-B_{\rm SDSS-EDR}=0.032\pm0.005$ mag with a standard deviation 
per galaxy of 0.094 mag and $B_{\rm MGC}-B_{\rm SDSS-DR1}=0.039\pm0.005$ mag 
with a standard deviation per galaxy of 0.086 mag. 

Since the 2dFGRS photometry and SCOS were taken from the same original UKST 
plates and the SDSS-DR1 is an update of the SDSS-EDR from the same CCDs, 
there are only 3 independent data sets. The best versions of these are the
MGC, SCOS and SDSS-DR1. We will concentrate
on the comparisons between these 3, with brief asides on the 2dFGRS and 
SDSS-EDR, since there are many publications that use photometry from these
datasets. From robust estimation via minimisation of the mean 
deviations (including $3-\sigma$ clipping) we determine the scale-errors 
between the different data sets. We fit the following equation and summarise 
the fits in Table~\ref{tab:phot}.

\begin{equation}
\Delta\,m  = a + b (B_{\rm MGC}-20)
\label{eq:scl}
\end{equation}

There is a $1.3$ per cent scale-error between the MGC and DR1 and a $4.5$ per 
cent scale error between the MGC and SCOS. However, there is only a $2.1$ per 
cent scale-error between SCOS and DR1. The reason these do not add up, is the
non-linearity in the scale-error between MGC and DR1 as can be seen from 
the large $\chi^2_{\nu}$ value. If we select objects over the same magnitude 
range,
$16<B_{\rm MGC}<19$, the 3 surveys become compatible, with the significant
change coming from an increased scale-error between MGC and DR1 ($2.7$ per 
cent). However, at the faint end, $19<B_{\rm MGC}<20$, the scale-error is
both small, $0.6$ per cent, and linear. The SDSS-EDR has the same scale-errors
as the DR1, but is fainter by $0.007$ mag and has a greater scatter 
($\sqrt{\sigma({\rm EDR})^2-\sigma({\rm DR1})^2}=0.04$ mag). The 2dFGRS has a 
very large scale error compared to the MGC (almost $6$ percent), but is only 
$1.2$ per cent different
from SCOS, which it was calibrated against. The scale errors are larger,
typically $2\pm1$ per cent at the bright end, $16<B_{\rm MGC}<19$, than the 
faint end, suggesting calibration problems associated with non-linearities, 
saturation or fewer standard stars. As we will show in Section 4.2, much of 
the variation is due to errors which are a function of surface brightness.  

The $3-\sigma$ clipped standard deviation (STD) of the overall magnitude
variance appears as expected between the MGC and 2dFGRS datasets ($\sigma_{\rm
STD} = \pm 0.14$ versus $\sigma_{\rm EXPECTED} = \pm 0.15$) but worse than
expected ($\sigma_{\rm STD} = \pm 0.09$ versus $\sigma_{\rm EXPECTED} = \pm
0.04$) between the MGC and SDSS-DR1. SCOS has a smaller variance w.r.t. the
MGC than 2dFGRS, while the SDSS-DR1 has a smaller variance than the SDSS-EDR,
demonstrating the improved photometry in both catalogues.

\begin{figure}
\centerline{\psfig{file=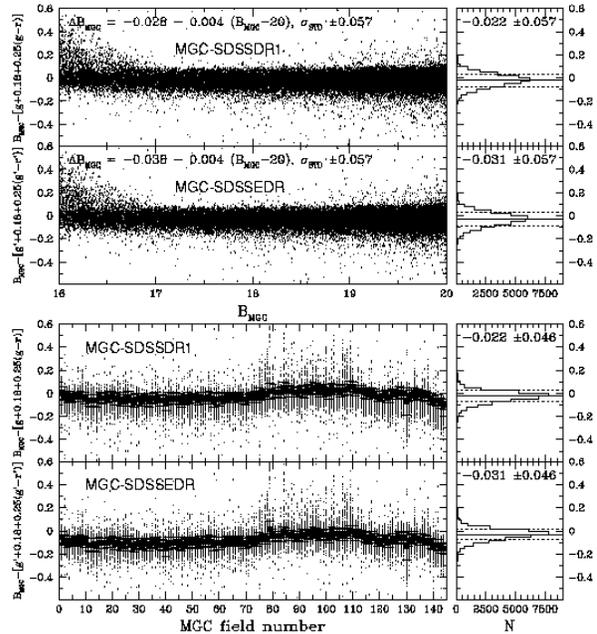,height=120.0mm,width=90.0mm}}
\caption{A comparison of the MGC and SDSS stellar photometry. The top 
panel shows the difference in photometry as a function of magnitude for 
the SDSS-DR1, where there is a $0.4$ per cent scale error, show by the best 
fit lines and a standard deviation of $0.057$ mag. The offset, $-0.022$ mag, 
appears different from the galaxy photometry, but when it is corrected for 
aperture differences it becomes $(0.044\pm0.005)$ mag, cf. $(0.039\pm0.005)$ 
mag. The similar plot for the SDSS-EDR is shown in the second from top plot. 
The only significant difference is in the offset, which is $-0.031$ mag. 
Table~\ref{tab:phstars} summarises the offsets
and fits to the scale error. In the 3rd and 4th panels we show the 
difference in photometry as a function of MGC field number, to determine the
uncertainty due to zeropoint errors across the MGC for the SDSS-DR1 and 
SDSS-EDR respectively. When we measure the mean standard deviation in each 
field we find that it is $0.046$ mag, suggesting that the MGC field offsets 
may be responsible for some fraction, ($\Delta ZP \sim \pm 0.035$ mag) of the 
general photometric discrepancy. The histograms to the right of the 3rd and
4th panels show the stellar population if the mean of each field is fixed to
be the mean of the whole distribution.}
\label{fig:offsets}
\end{figure}

To investigate whether this latter discrepancy may be due to systematic zero 
point (ZP) offsets between the individual MGC fields we show the equivalent 
trend for stars (Fig.~\ref{fig:offsets}, upper panels) and the ZP offset and 
standard deviation (Fig.~\ref{fig:offsets}, lower panels) per MGC field 
(using stars only). We then correct each individual magnitude 
by its respective field offset and rederive a ZP corrected $3-\sigma$ clipped 
mean for the full sample with a standard deviation $\pm 0.046$. This suggests 
that residual ZP offsets in the MGC {\it may} be at the level of up to 
$\pm 0.035$ mag, depending on variations across the SDSS, and therefore 
responsible for some fraction of this error. In Liske et al. (2003) we
find a standard deviation of $\pm 0.023$ mag in the offsets between adjacent
fields, with the most significant change occuring at field 74, as seen in 
Fig.~\ref{fig:offsets}. The scale error and variance 
error for stars is only improved marginally ($<0.001$ mag) between the 
SDSS-EDR and SDSS-DR1. The DR1 stellar magnitudes are $0.009$ brighter than 
those in the EDR, compared to $0.007$ mag brighter in the galaxy sample.
 
The larger variance for galaxies over stars suggests an additional
``galaxy-measurement'' error of $\pm 0.06$. This ``galaxy-measurement'' 
error consists in part of the increased signal-to-noise ratio per pixel
since galaxies are more extended than stars and also the expected variations 
between Kron and Petrosian magnitudes, which are anywhere from $+0.03$ mag 
for an exponential profile to $-0.12$ for a de Vaucouleur's profile. It is
difficult to calculate how large each component is, but it seems unlikely
that the variations between Kron and Petrosian would count for less than 
$\pm 0.03$, since this is the smallest expected variation for a particular 
galaxy, and could easily account for $\pm 0.04$ or $\pm 0.05$. This implies 
that improved consistency in galaxy photometry must come from a unified 
approach to galaxy photometry.

The offset in the stellar magnitudes between the MGC and SDSS-DR1 is 
$B_{\rm MGC}-[g+0.178+0.251(g-r)]=-0.022$ mag which is significantly 
different from the offset in the galaxy magnitudes between the 
MGC and SDSS-DR1, ($B_{\rm MGC}-[g+0.178+0.251(g-r)]=0.039$ mag. The stars 
can be approximated by a Gaussian profile, and the expected offset between 
the Kron and Petrosian magnitudes 
for Gaussian profiles is $m_{\rm Kron}-m_{\rm Pet}=-0.066$ mag. 
Thus the relative stellar magnitudes should be corrected by $0.066$ mag giving
a final value of $B_{\rm MGC}-[g+0.178+0.251(g-r)]=(0.044\pm0.005)$ mag with 
an individual scatter of $0.046$ mag for stars 
in the sample. The stellar and galaxy photometry agree to $0.005$ mag.

For SCOS, 2dFGRS and SDSS comparisons we also find a variation in the
variance as a function of magnitude as indicated, on 
Fig.~\ref{fig:phot_sing},  by the large 
solid data points (zero point offset per 0.5 mag) and error bars (one 
$3-\sigma$-clipped standard deviation). It is worth noting that the 2dFGRS 
shows larger photometric variance at brighter magnitudes whereas the SDSS 
shows increasing variance at faint magnitudes as one would expect for 
decreasing signal-to-noise data. SCOS shows both increasing variance at 
brighter and fainter magnitudes, with a minimum variance at
$B_{\rm MGC}\sim18.25$ mag. 

\subsection{Photometric variation with surface brightness}

\begin{figure}
\centerline{\psfig{file=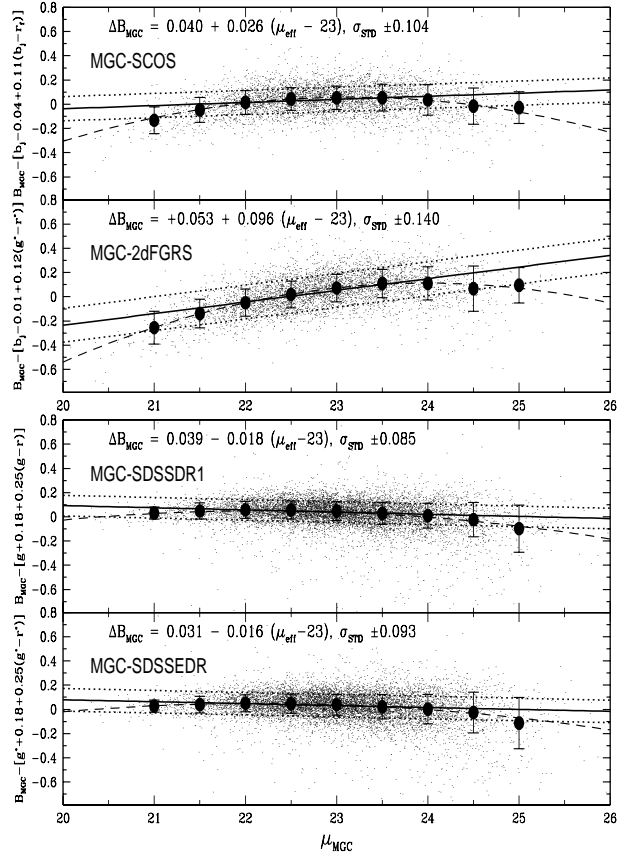,height=120.0mm,width=90.0mm}}
\caption{A comparison of the photometry between MGC and SCOS (top), MGC and 
2dFGRS (second from top), MGC and SDSS-DR1 (third from top) and MGC and 
SDSS-EDR (bottom), as a function of effective surface brightness. The best 
linear fit, via ($3-\sigma$) $\chi^2$ minimisation, is shown by the solid
lines. However, in each case a quadratic fit, shown by the dashed line,
gives a better fit. Table~\ref{tab:phtsb} gives the parameters for all the
fits. Each panel gives the best linear fit and standard deviation after 
removing this fit.}
\label{fig:phot_sb}
\end{figure}

Fig.~\ref{fig:phot_sb} shows the photometric variation as a function of 
effective surface brightness as defined in Section~\ref{sec:MGC}.

We fit the magnitude errors with the following equations:

\begin{eqnarray}
&\Delta\,m=a+b(\mu_{\rm eff}-23) \label{eq:sblin} \\ 
&\Delta\,m=\alpha+\beta(\mu_{\rm eff}-23)+\gamma(\mu_{\rm eff}-23)^2 \label{eq:sbqud}
\end{eqnarray}

None of the comparisons have good fits to Eqn~\ref{eq:sblin}, but almost
all have good fits to Eqn~\ref{eq:sbqud} indicating substantial 
non-linearities with surface brightness. Since the MGC is deeper than
SDSS, 2dFGRS and SCOS it is expected that both $\beta$ and $\gamma$ will
be small and negative, i.e. low surface brightness objects will be 
systematically fainter in the shallower surveys.

For the comparison between the MGC and the 2dFGRS we see a large positive
$\beta$ indicating a significant error in high surface brightness galaxies 
which galaxies at the $10^{th}$ percentile value of $\mu_{\rm eff}$, 
(21.5 mag arcsec$^{-2}$), offset from the mean by $-0.18$ mag, 
$B_{\rm MGC}-B_{\rm 2dFGRS}$. The low surface 
brightness galaxies have magnitudes closer to the mean value, with the 
largest offset ($+0.08$ mag) at the $90^{th}$ percentile value of 
$\mu_{\rm eff}$, (24.2 mag arcsec$^{-2}$). Since the high 
surface brightness galaxies are the most affected, the error is probably 
caused by non-linearities in the plates that have not been completely 
corrected during the calibration process. Any studies which utilise the 
2dFGRS photometry for structural analysis of the galaxy population 
(e.g. Cross et al. 2001) are thereby compromised. The SCOS data has a 
significant non-linearity too, $\gamma=-0.04$, but without the large 
linear error also evident in the 2dFGRS data. This results in an offset of 
$\sim-0.09$ mag at 21.5 mag arcsec$^{-2}$ and $\sim-0.02$ mag at 24.2 mag 
arcsec$^{-2}$.

In the MGC, SDSS comparison there is a small error at the low surface
brightness end, $\mu_{\rm eff}>24$ mag arcsec$^{-2}$. This error is 
$\sim-0.06$ mag difference from the EDR and $\sim-0.07$ mag difference in 
the  DR1 at the $90^{th}$ percentile value of 
$\mu_{\rm eff}$, (24.4 mag arcsec$^{-2}$). The error is as one might expect 
when comparing a deeper dataset with a shallower dataset and suggests that 
some flux is missing in the outskirts of low surface brightness galaxies in 
the SDSS-EDR data. While Kron and Petrosian magnitudes have little or no 
surface brightness dependency over a wide range of surface-brightness, 
inevitably they will miss flux from galaxies close to the detection threshold
since the profiles used to calculate the best aperture will be systematically
miscalculated at very low signal-to-noise ratios.

Comparisons between the other surveys indicate similar results, the deeper
survey finds more flux at the low surface brightness end, and that
there is a large $\sim10$ per cent error in the 2dFGRS with surface 
brightness and a smaller $\sim4$ per cent error in SCOS with surface
brightness. From Figs.~\ref{fig:comp_BBD} and ~\ref{fig:comp_BBD_s} it is 
clear that
bright galaxies are typically high surface brightness galaxies. The large
scale errors seen in the 2dFGRS and SCOS are due to these errors with 
surface brightness.

\subsection{Photometric accuracy of known low surface brightness galaxies}

As a slight digression we briefly 
address the specific question of the photometric accuracy
of low surface brightness galaxies. Impey et al. (1996) published a 
catalogue of luminous low surface brightness objects from stacked APM
scans in the equatorial region. From their full sample we note that 17
have positions inside the common MGC-2dFGRS-SDSS region. Of these
we find matches for all 17 from within the MGC and the SDSS, and for
15 within the 2dFGRS. One of the missing 2dFGRS objects was listed in Impey 
et al. as fainter than the 2dFGRS magnitude limit although both the MGC and 
SDSS magnitudes were above this limit. The other 2dFGRS failed match lay 
close to a bright star and is most likely a mis-classification or failed 
de-blend. 

\onecolumn

\tiny

\begin{table*}
\caption{A comparison of photometric matches between the three surveys and
the 17 low surface brightness objects from Impey et al. (1996; ISIB)
within the common region. Objects marked with $^1$ have multiple components in the SDSS. Each magnitude is the combination of all the components. All the
photometry is in $B_{\rm Johnson}$.}
\begin{tabular}{lrrrrrrrrrrr} \label{tab:isib}
ISIB ID   &RA(deg) & Dec(deg) & $B$ & $\mu_{\rm o}$& $\mu_{\rm eff}$ & MGC ID & $B_{\rm MGC}$  & $B_{\rm SCOS}$ & $B_{\rm 2dF}$& $B_{\rm EDR}$ & $B_{\rm DR1}$\\ \hline\hline
1035+0014 &159.606& $-$0.0183& 16.60& 22.40 & 24.60&  MGC90026& 16.44 & 16.78 & 16.49 & 17.06$^1$ & 16.74$^1$ \\
1042+0020 &161.287&  0.0753& 16.20& 21.40 &23.40&  MGC11548& 16.43 & 16.41& 16.48 & 16.67$^1$ & 16.69$^1$ \\
1043+0018 &161.561&  0.0500& 16.20& 21.40 &22.20&  MGC11695& 15.79 & 15.80& 15.85 &  15.88 & 15.93 \\
1045+0014 &162.082& $-$0.0225& 16.20& 21.80 &22.70&  MGC11884& 16.27 & 16.02& 15.95 & 16.39 & 16.21  \\
1102+0019 &166.166&  0.0572& 17.30& 24.10 &25.50&  MGC16030& 17.00 & 17.51& 17.47 & 17.83$^1$ & 17.39$^1$ \\
1125+0025 &172.123&  0.1436& 17.60& 23.00 &23.80&  MGC20736& 17.39 & 17.43& 17.11 & 17.54 & 17.52  \\
1129+0013 &172.994& $-$0.0508& 16.30& 22.30 &23.50&  MGC21656& 15.92 & 16.31& 15.85 & 16.15 & 16.20  \\
1216+0029 &184.858&  0.2150& 17.00& 22.90 &25.20&  MGC31502& 16.82 & 16.69& 16.63 & 17.73$^1$ & 17.82$^1$  \\
1221+0001 &185.926& $-$0.2575& 18.40& 23.90 &25.80&  MGC32646& 18.01 & 18.20& 18.20 & 18.47$^1$ & 18.63$^1$  \\
1221+0020 &186.127&  0.0708& 17.60& 23.70 &25.40&  MGC32544& 17.01 & 17.41&  17.22 & 17.34 & 17.45 \\
1247+0002 &192.519& $-$0.2339& 19.60& 26.40 &27.30&  MGC38179& 17.74 &  $---$   & $---$   & 18.16$^1$ & 19.17$^1$  \\
1310+0013 &198.187& $-$0.0406& 16.60& 22.30 &23.90&  MGC43127& 17.03 & 17.16& 16.86 & 18.36 & 18.34 \\
1405+0006 &212.132& $-$0.1272& 15.20& 20.60 &21.80&  MGC56580& 15.42 & 15.55 & 15.44 & 15.59$^1$ & 15.68$^1$ \\
1434+0020 &219.160&  0.1164& 17.50& 22.90 &24.50&  MGC64880& 17.21 & 17.22& 16.92 & 17.28 & 18.32 \\
1437+0001 &219.998& $-$0.1869& 18.50& 24.20 &25.40&  MGC66574& 18.02 & 18.36& 18.26 & 19.14 & 18.13 \\
1442+0026 &221.355&  0.2339& 16.70& 24.10 &24.30&  MGC90173& 16.44 & $---$  & $---$ & 16.51 & 16.53 \\
1158+0023 &180.159&  0.1106& 19.00& 24.40 &25.00&  MGC27147& 18.25 & 18.29& 18.31 & 19.30 & 18.54  \\ \hline

\end{tabular}
\end{table*}

\normalsize

\begin{figure}

\vspace{-7.0cm}

\centerline{\psfig{file=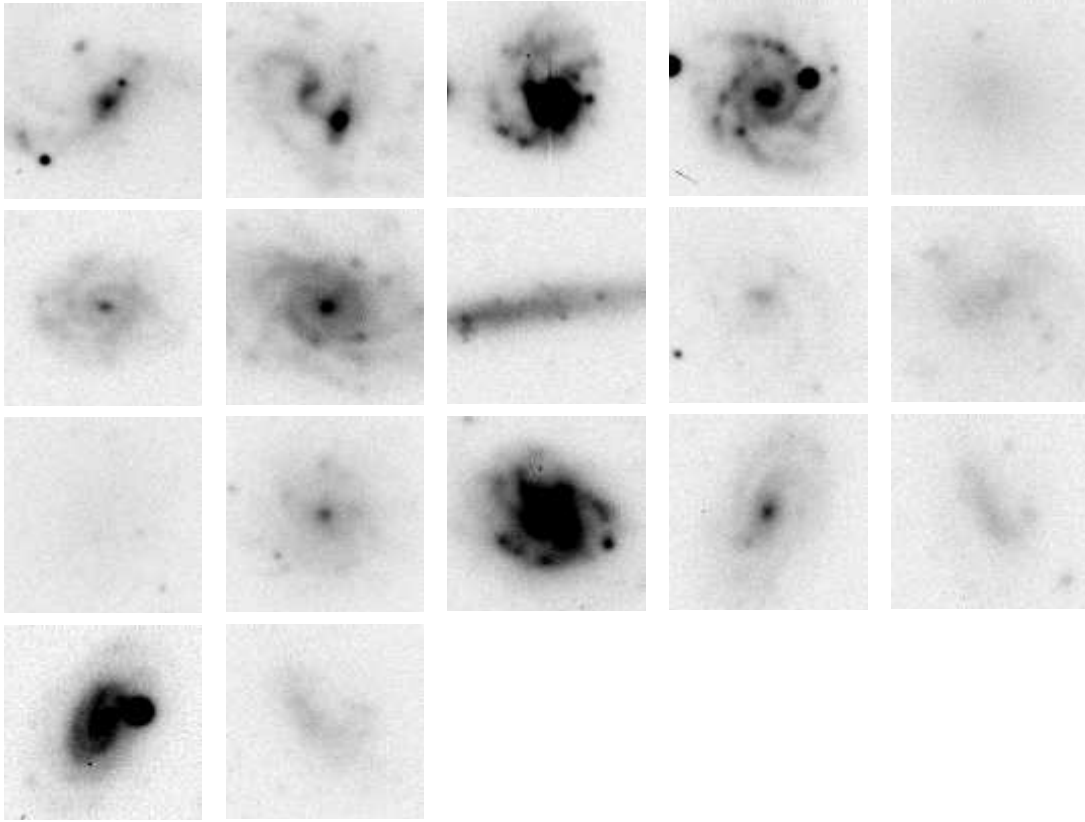,height=25.0cm,width=\textwidth}}

\vspace{-7.0cm}

\caption{The Impey et al. LSBGs in the MGC region. Each image is $33 \times 
33$ arcsec$^{2}$. The greyscale varies from 21 mag arcsec$^{-2}$ (black) to 
the $3\sigma$ 
variation in noise above the sky background (white).}
\label{fig:lsbgs}
\end{figure} 

\twocolumn

\begin{figure}
\centerline{\psfig{file=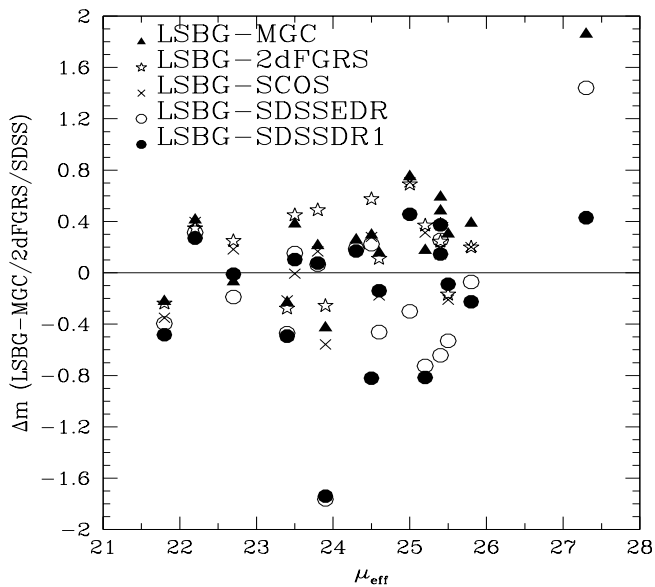,height=8.0cm,width=90.0mm}}
\caption{The photometric error between the surveys and the 
Impey et al. LSBGs in the common region as a function of effective
surface brightness.}
\label{fig:phot_lsbgs}
\end{figure}

The SDSS-EDR found multiple matches for 7 of the objects. Table~\ref{tab:isib}
shows the LSBG sample and the corresponding MGC matches for each survey,
along with the best $B_{\rm Johnson}$ magnitude for each of the data sets. 
Fig.~\ref{fig:lsbgs} shows a montage of these 17 objects from the MGC 
database. 
The magnitude zero point offset and $3-\sigma$ clipped standard deviations 
are: $B_{\rm LSBG}-B_{\rm MGC} = +0.22 \pm 0.32$, $B_{\rm LSBG}-
B_{\rm SCOS} = +0.07 \pm 0.32$, $B_{\rm LSBG}-B_{\rm 2dF} = +0.21 \pm 0.31$, 
$B_{\rm LSBG}-B_{\rm SDSS-EDR} = -0.17 \pm 0.65$ and 
$B_{\rm LSBG}-B_{\rm SDSS-DR1} = -0.16 \pm 0.57$ respectively. 
The MGC and 2dFGRS recover similar results, SCOS has a similar scatter but is 
0.2 mag fainter and both the SDSS-EDR and SDSS-DR1 are 0.2 mag fainter than
SCOS with greater scatter than the other 3. We also note from 
Fig.~\ref{fig:phot_sb} that the $\Delta (B_{\rm MGC} - B_{\rm SDSS})$ 
shows a larger dispersion than the $\Delta (B_{\rm MGC} - B_{\rm 2dF})$ in
the faintest surface brightness bin. This suggests that SDSS-EDR
photometry should be considered questionable for objects with 
$\mu_{\rm eff} > 24.5$ mag arcsec$^{-2}$.
Fig.~\ref{fig:phot_lsbgs} shows the $\Delta m$ versus $\mu_{\rm eff}$ 
derived from Table~\ref{tab:isib}
which clearly shows the degradation of photometric accuracy in the
SDSS data as a function of effective surface brightness
(filled and open circles representing SDSS-EDR and SDSS-DR1 data 
respectively). 
We also note the slightly upward trend in $\Delta (B_{\rm LSBG}-B_{\rm MGC})$ 
with increasing effective
surface brightness suggesting that the Impey et al. magnitudes themselves 
may be underestimating flux at the very low surface brightness end.
In particular the lowest low surface brightness galaxy (1247+0002), 
identified in both the MGC and SDSS-EDR, but not in the 2dFGRS, is 
considerably brighter in {\sc MGC-BRIGHT} and the SDSS-EDR than listed in 
Impey et al. (1996).
 
\section{Classification reliability}

All galaxies in {\sc MGC-BRIGHT} have been visually inspected and artefacts
reclassified, merged objects deblended and over-deblends reformed. The
MGC should therefore be considered robust. The 2dFGRS and SDSS-EDR
datasets use automated detection and classification algorithms over this 
magnitude range. It is therefore important to ascertain some independent
measure of the reliability of these large-scale surveys. Here we consider 
the accuracy of the automated classifiers in terms of star-galaxy 
separation, and contamination of the galaxy catalogues by stars or artefacts 
and galaxy incompleteness.

\subsection{Star-galaxy classification accuracy}

Although the 2dFGRS database is supposed to only include objects classified 
as galaxies, it is known to be contaminated by stars at the $5.4$ per cent 
level (cf. Norberg et al. 2001). 

We calculated the stellar contamination using the 2dFGRS-MGC catalogue.
There are 5241 good 2dFGRS objects matched to MGC objects\footnote{In a 
previous section we stated that there were 5285 2dFGRS objects matched to MGC 
objects. In that section we gave the number of objects for 
$[b_{\rm 2dF}-0.012+0.121(g^*-r^*)]>16$ mag, so that the number 
of non matches could be calculated. Here we give the number of objects for 
$B_{\rm MGC}>16$ mag, since the MGC is our yardstick.}.

Of these 368 are 
multiple matches, $(7.0\pm0.4)$ per cent, 178 are single stars and 4695 are 
single galaxies. The fraction of 2dFGRS mis-matches is not correlated with 
magnitude, as also noted in Norberg et al. (2002b). The fraction of single 
systems that are stars is $(3.7\pm0.3)$ per cent. The total fraction of 
2dFGRS objects containing stars is $(6.8\pm0.3)$ per cent and the fraction
in which the main component is a star is $(5.2\pm0.3)$ per cent. This  
agrees with the measurement of stellar contamination determined in earlier
papers and by the spectroscopic data (Colless et al. 2001). This 
indicates that while the star-galaxy separation algorithm does very well 
on individual stars and galaxies, it breaks down on close pairs.

The SDSS-EDR database includes stars and galaxies classified according to 
the criterion described in Stoughton et al. (2002). For our sample of 9795 
MGC galaxies ($16 < B_{\rm MGC} < 20$ mag), we find the following matches 
from the SDSS-EDR database 9656 galaxies, 20 stars and 119 non-detections. 
For our sample of 36260 MGC stars we find, in the SDSS-EDR database, 
305 galaxies and 35726 stars, leaving 229 non-detections. Since
we have spectra from various sources, it is possible to test the 
reliability of each classification. 

We can find the stellar contamination by dividing the number of objects 
classified as galaxies with $z<0.002$ by the total number of objects 
classified as galaxies. This measurement may be biased since the 
spectroscopic completeness varies with magnitude (see 
Section~\ref{sec:incom}).
We remove this bias by calculating this fraction as a function of magnitude
($f_{\rm st\,con}(B_{\rm MGC})$) and then multiplying by the total number 
of galaxies, to give the expected number of stellar contaminants at that 
magnitude. Furthermore we only use redshift data from the 2dFGRS and our own
redshift survey as these are only selected by magnitude and not colour. 

\begin{figure}
\centerline{\psfig{file=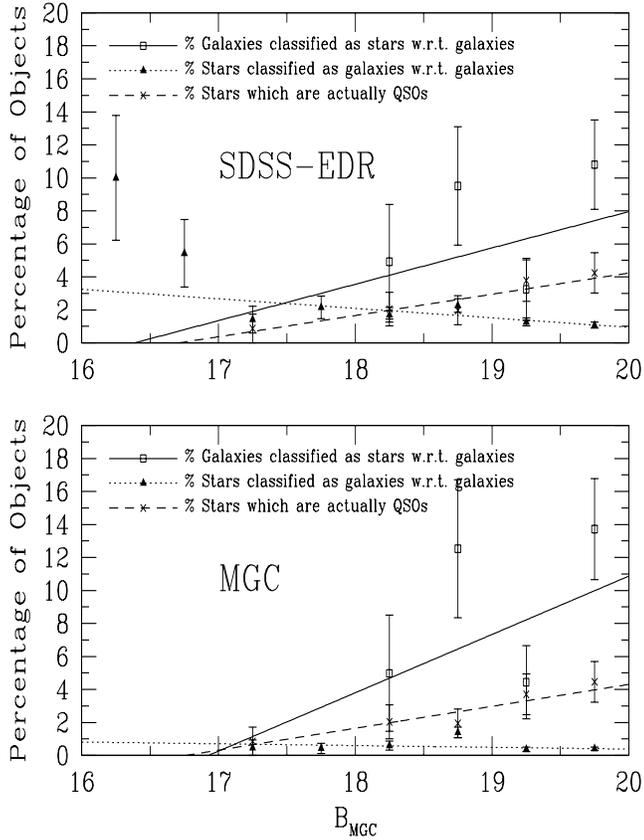,height=120mm,width=90.0mm}}
\caption{This plot shows misclassifications of stars and galaxies in the MGC
and SDSS-EDR as a function of magnitude. The triangles show the percentage
of stars contaminating the galaxy catalogue, as a function of magnitude. The 
dotted lines show the best linear fit to these data. The squares show the 
number of misclassified galaxies as a percentage of the galaxy catalogue, as 
a function of magnitude. The solid lines show the best
linear fit to these data. The crosses show the fraction of ``stars'' that
are in fact QSOs as a function of magnitude. The dashed lines show 
the best linear fit to these data.}
\label{fig:misclass}
\end{figure}

The fraction of stellar contamination at each magnitude is plotted 
as the triangles in Figure~\ref{fig:misclass}.
The best linear fit to the data is shown as the dotted line. The total 
stellar contamination is equal to the integral of this function over the
range $16 < B_{\rm MGC} < 20$ mag, see Eqn~\ref{eq:stcon}. In the MGC the
stellar contamination is ($0.47\pm0.07$) per cent and it is ($1.33\pm0.11$) per
cent in the SDSS-EDR.  

\begin{equation}
N_{\rm st\,con}=\int_{16}^{20}\frac{N_{\rm g,z<0.002}(m)}{N_{\rm g,all z}(m)}N_{\rm g}(m)\,dm
\label{eq:stcon}
\end{equation} 

\noindent where $N_{\rm g,z<0.002}(m)$ is the number of objects classified as 
galaxies with $z<0.002$, $N_{\rm g,all z}(m)$ is the number of objects 
classified as galaxies with a measured redshift and $N_{\rm g}(m)$ is the total
number of objects classified as galaxies all as a function of magnitude. 
$N_{\rm st\,con}$ is the total stellar contamination.

We perform similar calculations to find the number of galaxies misclassified
as stars w.r.t galaxies, see Eqn~\ref{eq:misclgl} and the number of stellar
objects that are QSOs, see Eqn~\ref{eq:qsos}. 

\begin{equation}
N_{\rm miscl\,gal}=\int_{16}^{20}\frac{N_{\rm s,0.002<z<0.4}(m)}{N_{\rm s,all z}(m)}\frac{N_{\rm s}(m)}{N_{\rm g}(m)}N_{\rm g}(m)\,dm
\label{eq:misclgl}
\end{equation} 

\noindent where $N_{\rm s,0.002<z<0.4}(m)$ is the number of objects 
classified as stars with $0.002<z<0.4$, $N_{\rm s,all z}(m)$ is the number of 
objects classified as stars with a measured redshift and $N_{\rm g}(m)$ is 
the total number of objects classified as galaxies, $N_{\rm s}(m)$ is the 
total number of objects classified as stars all as a function of 
magnitude and $N_{\rm miscl\,gal}$ is the total number of misclassified 
galaxies.

\begin{equation}
N_{\rm fr\,qso}=\int_{16}^{20}\frac{N_{\rm s,z>0.4}(m)}{N_{\rm s,all z}(m)}N_{\rm s}(m)\,dm
\label{eq:qsos}
\end{equation} 

\noindent where $N_{\rm s,z>0.4}(m)$ is the number of objects 
classified as stars with $z>0.4$, $N_{\rm s,all z}(m)$ is the number of 
objects classified as stars with a measured redshift and $N_{\rm s}(m)$ is 
the total number of objects classified as stars all as a function of 
magnitude. $N_{\rm miscl\,gal}$ is the total number of QSOs.

However, since the QSO
spectroscopic surveys targeting the stellar populations are mainly colour
selected, the objects from these surveys are not representative of the full
stellar population.  In our own redshift survey (MGCZ), we targeted the whole
population of stars and galaxies with $B_{\rm MGC}<20$ mag in some MGC
spectroscopic tiles. We use data from two such tiles (each tile is a separate
pointing of the 2dF instrument, and has a diameter of $1.95\deg$). The MGCZ
targets the remaining stellar targets once the data from other spectroscopic
surveys has been tallied, so it is important to use all the spectroscopic 
data available in these fields, not just MGCZ. This introduces some bias from 
the colour selected surveys if the
sample is not complete.  These tiles contain 1887 stars of which there are
1403 spectra from MGCZ and 53 spectra from other surveys. We find that $\sim 2$
per cent of these stars have redshifts of galaxies leading to a galaxy 
misclassification rate of ($6.6\pm1.3$) per cent in the MGC and ($5.3\pm1.0$) 
per cent in the SDSS-EDR. We estimate that $(5.6\pm1.3)$ percent of galaxies in the 
2dFGRS (galaxies in the MGC with $B_{\rm MGC}<19.0$) are misclassified as stars. 
The fraction of QSOs in the MGC stellar catalogue is ($2.1\pm0.4$) per cent and the 
fraction in the SDSS
stellar catalogue is ($2.2\pm0.4$) per cent. The effects of the bias from
colour selected surveys on the sample add an error of $\sim0.1$ per cent.

In each case the number of contaminants increases with magnitude. However, 
the fraction of stellar contamination of the galaxy catalogue  does not vary 
significantly with magnitude. The fraction of misclassified galaxies and the 
fraction of QSOs amongst the stars rise more steeply.

One caveat to the method above, is the cutoff redshift for stars and galaxies,
$z=0.002$. This was chosen based on the distribution of low redshift objects
in the 2dFGRS and assumes a Gaussian distribution of velocities for stars in 
the Milky Way. Since the Milky Way is a multi-component system, this limit
may miss some of the halo stars. While there are a few objects just above the
limit, which may turn out to be stars, this only reduces the numbers of 
misclassified galaxies by 22 per cent, from $(6.6\pm1.3)$ per cent to 
$5.1\pm1.3$ percent in the MGC. It still leaves a significant fraction of
misclassified galaxies. The misclassified galaxies will be discussed in more 
detail in a future paper (Liske et al. 2004).

\subsection{Artefacts}
\label{sec:art}

Of the 49 2dFGRS objects which were not matched to MGC objects,  
34 were 2dFGRS eyeball rejects i.e. inspections of the plates had already 
revealed them to be artefacts, 8 were not visible in Digital Sky Survey 
images (from the same Schmidt plates) and 7 looked like asteroids or 
satellite trails in the DSS images. None of these objects appeared in the 
SDSS. Thus all the extra 2dFGRS objects are accounted for and any objects missing
from the MGC are also missing in the 2dFGRS.

The SDSS-EDR contains 10213 galaxies and 34779 stars in the range 
($16 < B_{\rm MGC} < 20$ mag) of which 329 galaxies and 335 stars had no 
apparent counterparts in the MGC. These were checked by eye. They were missed
for various reasons (see Table~\ref{tab:art}). In some cases (292) the matching 
algorithm failed and in
another 99 cases the object was badly blended with a star, leading to
a disagreement in the deblending. There were another 6 faint smudges, near
bright objects. There were 119 'stars' and 66 galaxies seen in MGC images 
that do not appear in the MGC catalogues. This represents 0.35 percent of the 
stars and 0.67 percent of the galaxies. These missing objects are close
to bright stars and suggest that the exclusion regions are too conservative.

Finally, 46 galaxies and 36 stars had no counterparts in {\sc MGC-BRIGHT} or
{\sc MGC-FAINT} ($B_{\rm MGC} < 24$ mag) or flux in the images and are
therefore artefacts.  The proportion of artefacts appears to be 
$(0.45\pm0.07)$ per cent for galaxies and $(0.10\pm0.02)$ per cent stars. 
Table~\ref{tab:mcls} summarises the proportions of stellar contamination, 
misclassified galaxies and artefacts in each survey.

\begin{table}
\caption{A summary of the classification reliability of the 2dFGRS and 
SDSS-EDR galaxy catalogues from comparison with the MGC. Each number gives
the fractions as a percentage w.r.t. the galaxy population.}
\label{tab:mcls}
\begin{tabular}{lccc}
Catalogue & Stars classified & Gals classified & Artefacts \\ 
   &  as Galaxies  & as Stars   &  \\
\hline\hline
2dFGRS    & $ 5.2\pm0.3 $ & $5.6\pm1.3 $ & $ 0.9\pm0.1 $  \\
SDSS-EDR  & $ 1.3\pm0.1 $ & $5.3\pm1.0 $ & $ 0.45\pm0.07 $ \\
MGC       & $ 0.47\pm0.07 $ & $6.6\pm1.3 $ & --- \\
\hline
\end{tabular}
\end{table}

\begin{table}
\caption{A summary of objects in the SDSS-EDR with no counterpart 
in the MGC.}
\label{tab:art}
\begin{tabular}{lcccc}
Reason & No of  & No of & $\%$-age of & $\%$-age of \\
 & Stars & Galaxies & Stars & Galaxies \\
\hline\hline
Failed match & 101 & 191 & 0.29 & 1.9 \\
Badly blended & 79 & 20 & 0.23 & 0.20 \\
Smudges near & 0 & 6 & 0. & 0.06  \\
bright objects & & \\
Detected but  & 119 & 66 & 0.34 & 0.65 \\
not catalogued & & \\
Artefacts & 36 & 46 & 0.10 & 0.45 \\
\hline
Total & 335 & 329 & 0.96 & 3.2  \\
\hline
\end{tabular}
\end{table}

\section{Incompleteness}
\label{sec:incom}

The magnitude limit of the 2dFGRS catalogue is nominally 
$b_{\rm 2dF,old,lim}=19.45$
mag. However, the photometry of objects in the 2dFGRS has been revised since 
the target catalogues were produced, so there is not a single magnitude limit.
Two of the plates (UKST 853 and 866) have particularly bright limiting 
magnitudes, $b_{\rm 2dF,lim}<19.18$ mag, so we have removed these plates when 
testing the completeness. The plates are removed by selecting MGC galaxies in 
the range $153.145\deg<{\rm RA}<213.145\deg$ and ${\rm RA}>218.145\deg$. 
Galaxies in this part of the 2dFGRS have 
$b_{\rm 2dF,lim} \ge 19.23$ mag. This corresponds to 
$B_{\rm MGC}=19.365$ mag. If we test the completeness at $B_{\rm MGC}=19.0$ 
mag, 2.56 standard deviations brighter than this limit, only $0.5$ per cent 
of the 2dFGRS data at this magnitude (i.e. $<5$ objects) will be missing 
due to random errors.

\begin{figure}
\centerline{\psfig{file=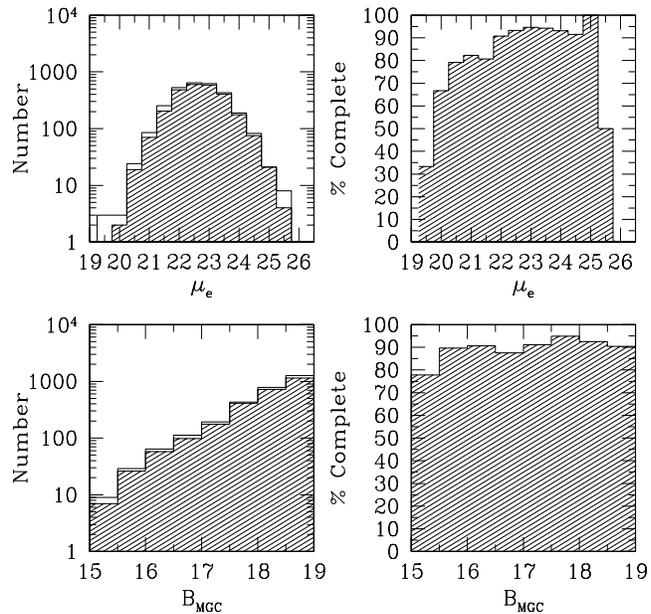,width=85mm,height=90mm}}
\caption[Incompleteness of 2dFGRS vs $B_{\rm MGC}$]{The variation of the 
incompleteness of the 2dFGRS with $B_{\rm MGC}$ (lower) and $\mu_{\rm eff}$ (upper). 
The left-hand plots show the histogram of the total number of galaxies in each 
bin (solid line), and the histogram of the number with redshifts (filled). 
The right hand side histogram shows the completeness percentage in each bin.}
\label{fig:compm}
\end{figure}

There are 2891 MGC galaxies in the correct RA range with 
$B_{\rm MGC}<19.0$ mag. This catalogue was separated into objects with a 
2dFGRS match and objects 
without. MGC objects that are a member of a multiple system of 2 or more MGC
objects matched to a single 2dFGRS object were placed into the matched bin if 
they were the principle component and into the non-matched bin if they were a 
secondary component. There were
2646 matches, giving a completeness rate of $(91.3\pm1.8)$ per cent. The 
variation of incompleteness with magnitude is shown in Fig.~\ref{fig:compm}.
The variation is consistent with a constant incompleteness, so the incompleteness 
at $b_{\rm 2dF}=19.45$ mag is $IC_{\rm b_j 19.45}=(8.7\pm0.6)$ per cent. 
This result matches the result from Norberg et al. (2002b) which gives a value
$IC_{\rm b_j 19.}=(9\pm2)$ per cent. It is marginally greater than the original
APM expectation of $3-7$ per cent incompleteness.

The variation with effective surface brightness is also shown in
Fig.~\ref{fig:compm}. For $22.5<\mu_{\rm eff}<24.5$ mag arcsec$^{-2}$ the
completeness is fairly constant $IC\sim5$ per cent. The incompleteness of
LSBGs increases rapidly beyond $\mu_{\rm eff}=25.0$ mag arcsec$^{-2}$ and no
2dFGRS galaxies are seen with $\mu_{\rm eff}>25.75$ mag arcsec$^{-2}$, as
expected with an isophotal limit $\mu_{\rm b_j,lim}=24.67$ mag arcsec$^{-2}$
(see Cross et al. 2001) and an exponential profile ($\mu_{\rm
eff}=\mu_0+1.124$ mag arcsec$^{-2}$). At the bright end, the incompleteness
rises steadily. Since the incompleteness rises steadily for faint objects and
also for high-surface brightness objects, it would make sense if a significant
proportion of the missing objects are both faint and high-surface brightness
i.e., compact galaxies which looked like stars on the Schmidt plates.

The variation with surface brightness is consistent with the Norberg et
al. (2002b) comparison between the 2dFGRS and SDSS-EDR, given that the peak
in the surface brightness distribution of his sample is $\mu_{\rm b_j}=22.2$
mag arcsec$^{-2}$ and the peak in the surface brightness distribution of our
sample is $\mu_{\rm e,B_{\rm MGC}}=22.9$ mag arcsec$^{-2}$.  At the high surface
brightness end, we both measure a decrease in completeness at $\mu_{\rm
e,B_{\rm MGC}}\sim21.7$ mag arcsec$^{-2}$ ($\mu_{\rm b_j}\sim21.0$ mag
arcsec$^{-2}$). At the low surface brightness end, the decline in completeness
appears to occur at a slightly different point, but is consistent with the
errors and the incompleteness of the SDSS, see Section~\ref{sec:incomSDSS}.

\subsection{Types of galaxy missing from 2dFGRS}
\label{sub:typ_miss}

\begin{figure}
\centerline{\psfig{file=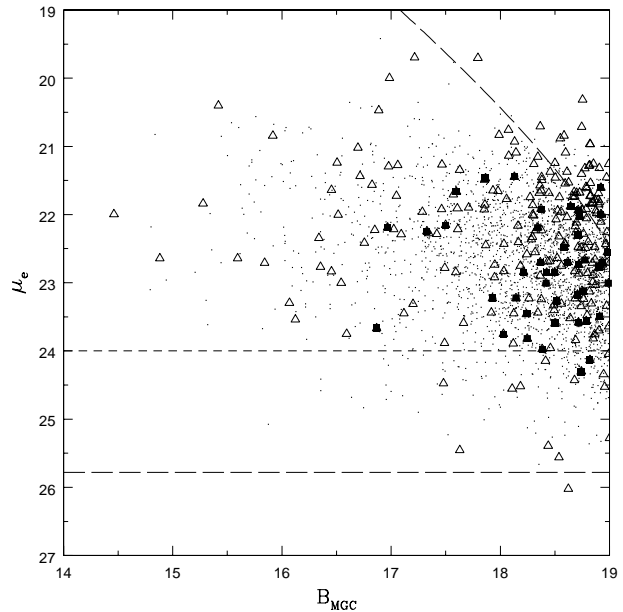,width=85mm,height=85mm}}
\caption[Apparent BBD showing missing 2dFGRS galaxies]{Plot of all galaxies 
$B_{\rm MGC}\leq\,19.0$ in the $B_{\rm MGC},\mu_{\rm eff}$ plane. The dots 
represent galaxies 
with 2dFGRS matches, the triangles represent those without matches and the 
squares represent those without matches which are the secondary components 
of 2dFGRS matches. The lower horizontal line represents the theoretical limit 
at which 2dFGRS galaxies can be seen. The upper horizontal line represents 
the limit at which a significant fraction of 2dFGRS galaxies are being missed.
Objects below this line are probably missed because of their low surface 
brightness. The curved line represents an exponential galaxy with a radius of 
$3.6\arcsec$. More compact objects were excluded because they were classified 
as stars - see Cross et al. (2001).}
\label{fig:comp_BBD}
\end{figure}

Fig.~\ref{fig:comp_BBD} shows all the galaxies $B_{\rm MGC}\leq\,19.0$ mag
plotted in the $B_{\rm MGC},\mu_{\rm eff}$ plane. The lower horizontal line 
represents the limit at which low surface brightness galaxies (LSBG) would be 
expected to be missed from the 2dFGRS. The upper horizontal line represents
the limit at which a galaxy is classified as a LSBG ($\mu_{\rm eff}>24.0$ mag
arcsec$^{-2}$). These galaxies make up $6.1$ per cent of the 
population of missing galaxies. The curved line represents the 
rough star-galaxy separation line. This curve is the locus of disk galaxies 
with $r_{\rm iso}=3.6\arcsec$ when $\mu_{\rm b_j,lim}=24.67$ mag 
arcsec$^{-2}$. While the APM detects objects with a minimum of 16 pixels, 
corresponding to 4 arcsec$^{2}$, the histogram of objects in the 2dFGRS has a
minimum radius of $\sim3.6\arcsec$. The $17.8$ per cent of objects faintwards 
of this line are more compact than this and are likely to be unresolved. 
The squares represent 
objects which are secondary components of 2dFGRS galaxies, missed because of 
poor deblending. These make up $17.4$ per cent of missing galaxies, although 2 
galaxies are missed because they are thought to be stars with poor
deblending. These objects account for $39.7$ per cent of missing objects. 

We looked for the other 60 per cent of missing 2dFGRS objects in the full 
APM catalogue. 
The APM catalogue contains many objects that did not make the final 2dFGRS
selection catalogue due to difficulties getting spectra: e.g. other nearby 
galaxies or stars. The rest of the missing 2dFGRS objects were compared to 
these objects. The excluded objects included blended objects, unresolved 
objects and some normal galaxies. After looking at these objects, it was 
discovered that $(53\pm5)$ per cent of all missing 2dFGRS objects were 
classed as blended, or were secondary objects matched to a 2dFGRS object, 
$(18\pm3)$ per cent were unresolved, $(19\pm3)$ per cent were normal 
galaxies and $(6\pm2)$ per cent were LSBGs.

Blended objects are those resolved by the APM, which were still too close
together for the 2dF spectrograph to be able to adequately handle. Secondary
objects were those too close to another object to be resolved by the APM.

Pimbblet et al. (2001) have also looked at the completeness of the APM by 
matching it to Las Campanas / AAT Rich Cluster Survey (LARCS) data for 4 
Abell clusters. They find a higher overall incompleteness rate, with 
$10-20$ per cent of galaxies missing at all magnitudes, 
$b_{\rm 2dF}\leq\,18.85$ mag, and $\sim20$ per cent missing for 
$b_{\rm 2dF}<17.0$ mag. The denser environment of clusters might explain why a 
larger fraction of objects are missing in the LARCS data since one would
expect more blends. However, Pimbblet et al. (2001) show that there
is no increase in total fraction or blended fraction close to the cluster 
centres.

They find that $60$ per cent of missing objects are blends, $15$ per 
cent were unresolved galaxies, $20$ per cent are normal galaxies and 
$5$ per cent are LSBGS. Pimbblet et al. also find the median merger distance 
for blends, which varies from $(5.3\pm0.9) \arcsec$ in Abell 1084 to 
$(8.6\pm0.9) \arcsec$ in Abell 22.  

The LARCS group also determined why galaxies have been missed in the 
2dFGRS. Missing blended, unresolved objects, and LSBGs are easily understood, 
but it is difficult to comprehend why those galaxies classified as normal 
are missing. Pimbblet et al. found that these objects had been classified 
as ``stellar'', ``blended'' or ``noise'' on APM R-band plates which were
used jointly with the $b_{\rm j}$ plates to classify objects. The original APM
catalogues are complete for all galaxies apart from some LSBGs, secondary 
galaxies and poorly resolved galaxies (about $3.7$ per cent of all galaxies, 
$B_{\rm MGC}<19.0$ mag) but the 2dFGRS target catalogue is less complete, 
missing $(8.7\pm0.6)$ per cent of $B_{\rm MGC}<19.0$ mag galaxies.

Finally Pimbblet et al. (2001) showed that the proportion of blends and 
unresolved galaxies missing in the 2dFGRS is constant with magnitude, 
whereas normal and low-surface brightness galaxies are missed predominantly 
at $b_{\rm 2dF}>18.0$ mag. A modest increase in incompleteness is seen for 
fainter galaxies, but the uncertainties are such that the results are 
consistent with constant incompleteness.

\subsection{Incompleteness of the SDSS-EDR}
\label{sec:incomSDSS}

\begin{figure}
\centerline{\psfig{file=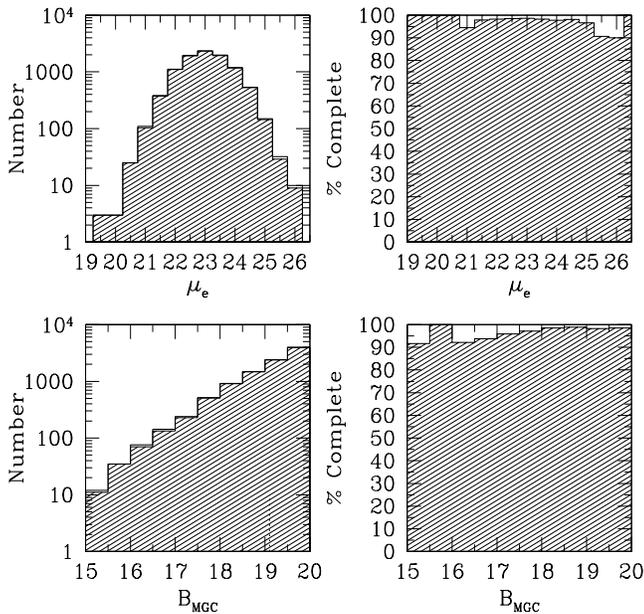,width=85mm,height=90mm}}
\caption[Incompleteness of SDSS vs $B_{\rm MGC}$]{The variation of the 
incompleteness of the SDSS-EDR with $B_{\rm MGC}$ (lower) and $\mu_{\rm eff}$ (upper). 
The left-hand plots show the histogram of the total number of galaxies in each 
bin (solid line), and the histogram of the number with redshifts (filled). 
The right hand side histogram shows the completeness fraction in each bin.}
\label{fig:inc_s}
\end{figure}

We have also checked the incompleteness of the SDSS-EDR. Out of 9795 MGC
galaxies ($B_{\rm MGC}<20.$) mag, the overall incompleteness is $(1.8\pm0.1)$ 
per cent.
Fig.~\ref{fig:inc_s} shows the photometric incompleteness of the SDSS-EDR 
as a function of $B_{\rm MGC}$. The incompleteness is never greater than $3$ 
per cent at any magnitude. 

Fig.~\ref{fig:inc_s} shows the photometric incompleteness of the SDSS-EDR 
as a function of $\mu_{\rm eff}$. The incompleteness is $\leq5$ per cent for 
$21.5<\mu_{\rm eff}<25.0$ mag arcsec$^{-2}$. It rises when 
$\mu_{\rm eff}>25.0$ mag arcsec$^{-2}$ due to the low signal to noise of these
galaxies in the SDSS. While only $(2.0\pm0.4)$ per cent of LSBGs 
$\mu_{\rm eff}>24.0$ mag arcsec$^{-2}$ are actually missing, this represents
$(13.5\pm2.8)$ per cent of all missing SDSS galaxies.  

\begin{figure}
\centerline{\psfig{file=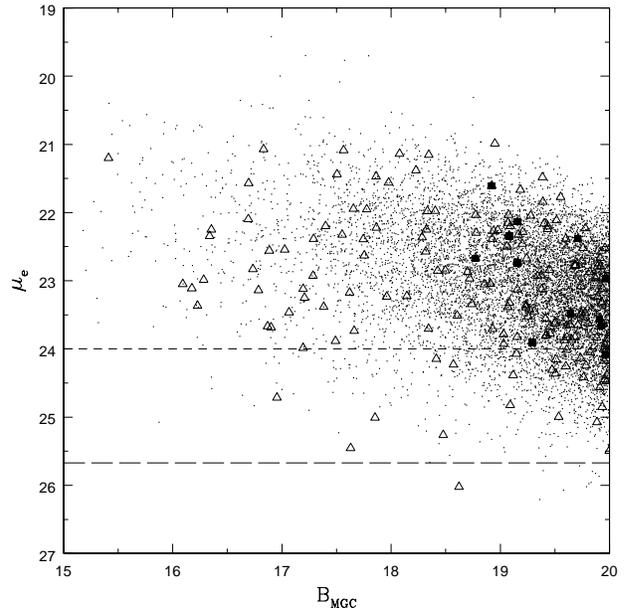,width=85mm,height=85mm}}
\caption[Apparent BBD showing missing 2dFGRS galaxies]{Plot of all galaxies 
$B_{\rm MGC}\leq\,20.0$ in the $B_{\rm MGC},\mu_{\rm eff}$ plane. The dots represent galaxies 
with SDSS-EDR matches, the triangles represent those without matches and the 
squares represent those without matches which are the secondary components 
of SDSS-EDR matches. The lower horizontal line represents the theoretical 
limit at which SDSS-EDR galaxies can be seen. }
\label{fig:comp_BBD_s}
\end{figure}

Fig.~\ref{fig:comp_BBD_s} shows the distribution of missing galaxies 
(triangles) as a function of $B_{\rm MGC}$ and $\mu_{\rm eff}$. The squares 
show missing galaxies, where the MGC galaxy is the secondary component of an 
SDSS galaxy ($32.8\pm4.3$ per cent of cases).

\subsection{Magnitude and surface brightness biases in incompleteness}

It is important to select a region of parameter space with high completeness 
when measuring the space density. If a region has high photometric 
incompleteness, then many objects have been missed from the input catalogues, 
e.g. compact objects that are thought to be stars, LSBGs. We have no 
information about these objects and can only speculate on their importance to
the overall luminosity and mass density. In regions where the photometric
incompleteness is high, then the redshift incompleteness will also be high, 
but there can be additional regions where the photometric incompleteness is
low, but the redshift incompleteness is high. This may be for a variety of 
reasons: low signal to noise in the spectrograph, or objects which are 
only found in clusters may be missed preferentially because of the minimum 
separation of fibres. 

Thus the calculation of the space density will only be robust in regions 
where both the photometric and the redshift completeness are high. In regions 
where the redshift completeness is low, the question is: have the missing 
objects got the same redshift distribution as those objects with redshifts? 
This may be a plausible 
assumption, but objects with spectral lines close to sky lines may be missed
preferentially, or objects with weak emission or absorption may be missed in 
preference to those objects with strong lines. If the photometric 
incompleteness is
high, not only do we have these problems, but we also have to wonder if there
is redshift or other bias in the missing objects. As shown in 
Section~\ref{sub:typ_miss}, there are many blended objects and compact 
objects missing from the 2dFGRS. 
These may be preferentially missed from cluster environments where a lot of
galaxies have a similar redshift. Thus the redshift distribution seen in that
region of parameter space may be less clustered than the true redshift 
distribution.

\begin{figure}
\centerline{\psfig{file=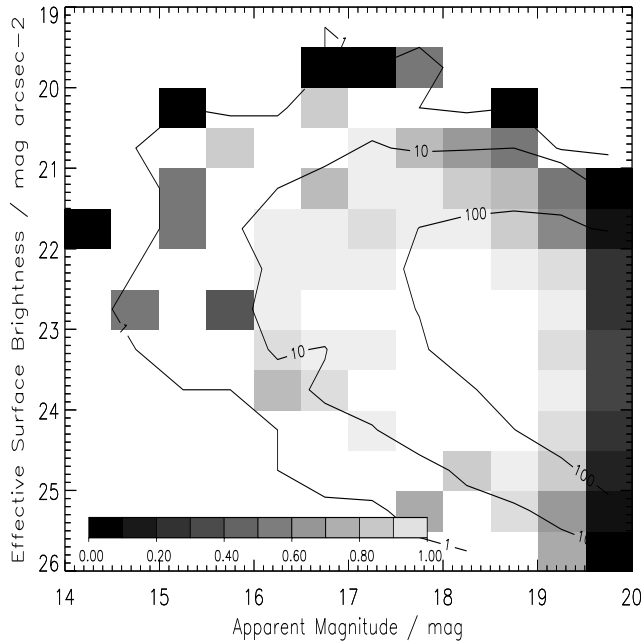,width=90mm,height=90mm}}
\caption[Photometric completeness of 2dFGRS]{This figure shows 
the photometric completeness of the 2dFGRS imaging catalogue as a function of 
$B_{\rm MGC}$ and $\mu_{\rm eff}$. The grey-scale represents the 
completeness fraction of galaxies. The contours represent the total 
number of MGC galaxies in each bin. Outside of the $N_{\rm tot}=1$ line 
there is no data.
\label{fig:comp_ph_TdF}}
\end{figure}

\begin{figure}
\centerline{\psfig{file=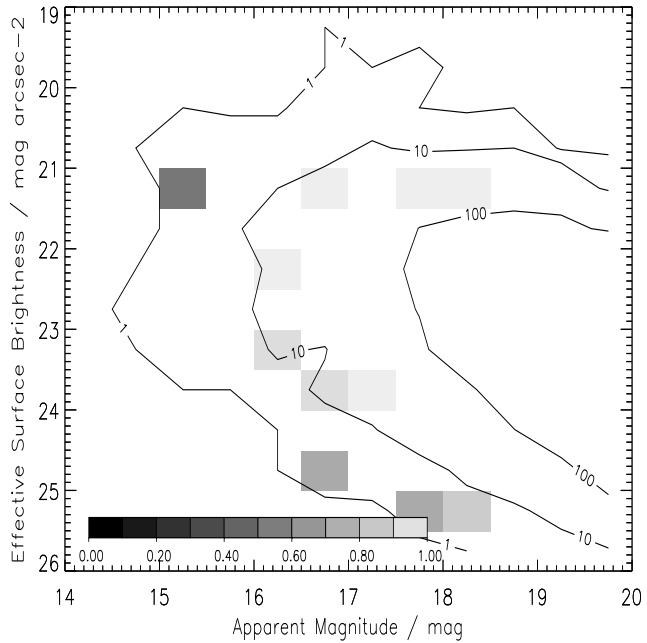,width=90mm,height=90mm}}
\caption[Photometric completeness of SDSS-EDR]{This figure shows 
the photometric completeness of the SDSS-EDR imaging catalogue as a function 
of $B_{\rm MGC}$ and $\mu_{\rm eff}$. The grey-scale represents the 
completeness fraction of 
galaxies. The contours represent the total number of MGC galaxies in each 
bin. Outside of the $N_{\rm tot}=1$ line there is no data.
\label{fig:comp_ph_SDSS}}
\end{figure}

Fig.~\ref{fig:comp_ph_TdF} shows the photometric completeness of the 
combined 2dFGRS dataset as a function of both magnitude and surface 
brightness. Fig.~\ref{fig:comp_ph_SDSS} shows the equivalent plot for the 
SDSS-EDR. In the case of the 2dFGRS, the completeness is very low ($<40$ per 
cent) for $B_{\rm MGC}>19.5$ and also very low for faint high surface 
brightness galaxies, that may be confused with stars. The SDSS-EDR on the 
other hand has very high completeness ($>90$ per cent) in virtually every bin.

\begin{figure}
\centerline{\psfig{file=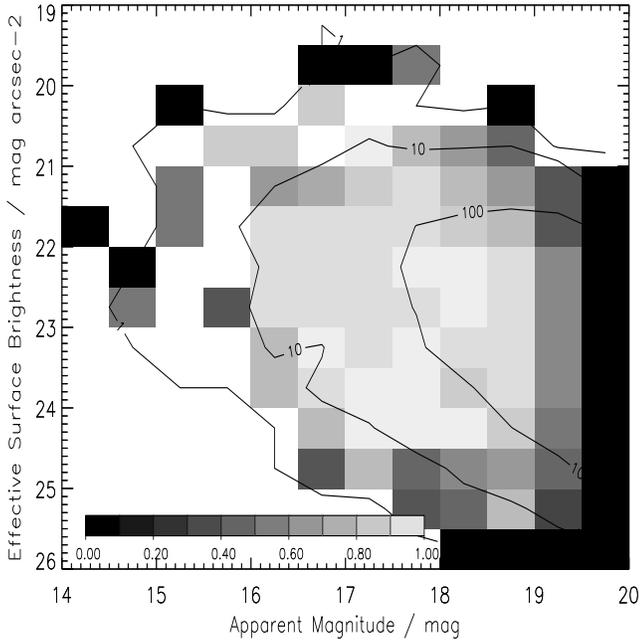,width=90mm,height=90mm}}
\caption[Redshift Completeness of 2dFGRS]{This figure shows the 
redshift completeness of the 2dFGRS imaging catalogue as a function of 
$B_{\rm MGC}$ and $\mu_{\rm eff}$. The grey-scale represents the 
completeness fraction of galaxies. The 
contours represent the total number of MGC galaxies in each bin. Outside of 
the $N_{\rm tot}=1$ line there is no data.
\label{fig:comp_z_2dF}}
\end{figure}

\begin{figure}
\centerline{\psfig{file=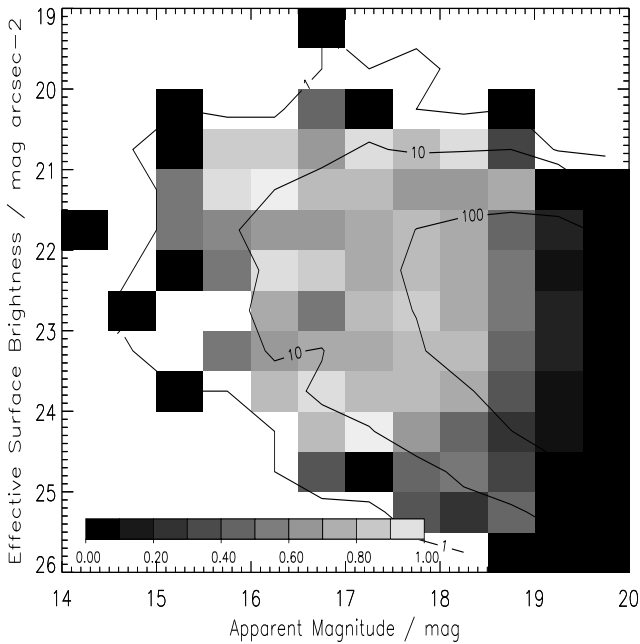,width=90mm,height=90mm}}
\caption[Overall Redshift Completeness of SDSS-EDR]{This figure
 shows the redshift completeness of the SDSS-EDR imaging catalogue as a 
function of $B_{\rm MGC}$ and $\mu_{\rm eff}$. The grey-scale represents the completeness 
fraction of galaxies. The contours represent the total number of MGC galaxies 
in each bin. Outside of the $N_{\rm tot}=1$ line there is no data.
\label{fig:comp_z_SDSS}}
\end{figure}

Fig.~\ref{fig:comp_z_2dF} shows the redshift completeness of the 2dFGRS. 
Fig.~\ref{fig:comp_z_SDSS} shows the redshift completeness of the SDSS-EDR. 
For the 2dFGRS the original spectroscopic magnitude limit was 
$b_{\rm 2dF}=19.45$ mag, 
but this has now become a variable with plate number and dust correction. 
The analysis above showed that the limit for high completeness is 
$B_{\rm MGC}\sim19.0$ mag. For the SDSS-EDR, the spectroscopic limit is 
$r^*=17.7$ mag for most galaxies. The filter conversion equation is 
$B=g^*+0.251(g^*-r^*)+0.178$ which converts to 
$B_{\rm MGC}=r^*+1.251(g^*-r^*)+0.178$. Using a typical $(g^*-r^*)=0.6$, 
$B_{\rm MGC,lim}=18.7$ mag for the SDSS-EDR. However, the completeness may 
drop before this limit or after this limit due to the variation in colours of 
galaxies in the sample. It is apparent that the redshift completeness of 
the SDSS-EDR falls off $\sim0.5$ mag brighter than the 2dFGRS. The 2dFGRS is 
more complete than the SDSS-EDR because it has been finished whereas there 
are some small gaps in the SDSS-EDR spectroscopic release. In the MGC these
gaps occur at $152.7<{\rm R.A.}<153.4$, $153.9<{\rm R.A.}<155.5$, 
$168.5<{\rm R.A.}<170.5$, $203.0<{\rm R.A.}<204.8$. The spectroscopic sample 
of the SDSS-EDR was selected
in the $r^*$ filter, so bluer galaxies will have a brighter $B_{\rm MGC,lim}$
and redder galaxies will have a fainter $B_{\rm MGC,lim}$. It is not the 
overall redshift completeness that concerns us, but rather how the 
completeness varies with magnitude and surface brightness.

There is a reduced level of redshift completeness in both surveys for 
galaxies with 
$\mu_{\rm eff}>24.5$ mag arcsec$^{-2}$, whereas the photometric completeness
dropped most significantly for objects with 
$\mu_{\rm eff}>25.0$ mag arcsec$^{-2}$. 

\section{Conclusions}

In this paper we used a deep wide field CCD imaging survey, the MGC 
(Liske et al. 2003) to test the photometric accuracy and completeness 
of the 2dFGRS and SDSS-EDR, as well as the photometric accuracy of SCOS
and SDSS-DR1.

The main photometric and completeness results for the 2dFGRS and SDSS 
are summarised in Tables~\ref{tab:phot}, ~\ref{tab:phstars}, ~\ref{tab:phtsb},
~\ref{tab:mcls}, and ~\ref{tab:art}. The comparison between the MGC and 
SDSS-DR1 finds that 
$\Delta\,m=0.039\pm0.005$ mag with a scatter of 0.086 mag per galaxy. The 
stellar catalogue has photometry with $\Delta\,m=0.044\pm0.005$ mag with a 
scatter of 0.046 mag, once the difference between Kron and Petrosian 
magnitudes and the field-to-field scatter has been taken into account. The 
field-to-field scatter in the MGC is $\sim0.035$ mag. We estimate that the ``Galaxy Measurement'' 
error, a combination of decreasing signal-to-noise per pixel and the 
differences between Kron and Petrosian magnitudes contribute to a scatter 
of 0.06 magnitudes in the galaxy errors. There is a small scale error of
$2.7$ per cent for bright galaxies, but faint galaxies and stars have 
extremely small scale errors of $\sim0.5$ per cent compared with the MGC.
However, the fluxes of low surface brightness galaxies 
$\mu_{\rm eff}>24.5$ mag arcsec$^{-2}$ are systematically 
underestimated by $\sim0.1$ mag. 

The SDSS-EDR has similar scale-errors and errors with surface brightness
to SDSS-DR1. The significant differences, are an offset of $0.007$ 
magnitudes, with the SDSS-DR1 magnitudes slightly brighter and also
a reduced standard deviation per galaxy for the SDSS-DR1. 

Since the 2dFGRS and SCOS magnitudes in the 2dFGRS database have been 
calibrated using SDSS-EDR photometry, the offset w.r.t. the MGC should be 
the same and indeed it is within the expected errors. Of the four data sets 
compared to the MGC, the 2dFGRS has the worst photometry with 
$B_{\rm MGC}-B_{\rm 2dF}=(0.035\pm0.005)$ mag with a scatter of 0.142 mag 
per galaxy and a very large 
scale error, $5.7$ per cent which probably comes from non-linearities in the 
photometric plates causing the flux of high-surface brightness objects
to be significantly underestimated (see Fig~\ref{fig:phot_sb}). High surface 
brightness objects have their fluxes underestimated in the 2dFGRS by 
$\sim0.18$ mag.  

The SCOS magnitudes are a significant improvement on the
2dFGRS magnitudes, with a lower variance and especially with regard to the 
variation in $\Delta\,m$ with surface brightness. This results in a reduced 
scale error and the SCOS photometry is well matched to the SDSS-EDR. However, 
while the scale-error compared to the MGC is lower than the 2dFGRS 
scale-error, it is still quite large ($4.5$ per cent). Both the SCOS and
SDSS-DR1 show significant improvements when compared to the 2dFDRS and 
SDSS-EDR respectively, as expected with later releases.

The main source of error in the comparison of the offsets is the colour
equations used to compare the photometry. These need to be accurate to 
$<0.002$ magnitudes before random errors become the main source of error in
the comparison between 2dFGRS and MGC.

While it is impossible to say for certain which survey has the best 
photometry, as all the checks are relative, some trends can be seen. The 
MGC seems to be fainter by 0.04 mag than all the other surveys, but since the
2dFGRS and SCOS are matched to the SDSS-EDR, we have to be very careful on
this matter. The scale-error results cannot be interpreted in an absolute 
sense either. However the standard deviation per galaxy can be a useful 
indicator. It is lowest between the MGC and DR1 (apart from between the EDR 
and DR1 which are taken from the same data) indicating that these are the two 
best surveys. It is difficult to tell if
the MGC or the DR1 is the best since the MGC has the lowest standard deviation
compared to the APM, but DR1 has the lowest compared to SCOS. 

We find that $(5.2\pm0.3)$ per cent of the objects classified as galaxies in 
the 2dFGRS are stars, $(7.0\pm0.4)$ per cent are multiple objects and 
$(0.9\pm0.1)$ per 
cent artefacts. When compared to the MGC galaxy catalogue, we find that the
2dFGRS is incomplete by $(8.7\pm0.6)$ per cent by $B_{\rm MGC}=19.0$ mag. Since the
spectroscopic data show that $(5.6\pm1.3)$ per cent of the MGC galaxy 
catalogue are misclassified as stars then we conclude that $(14.3\pm1.4)$ per 
cent of galaxies are missing from the 2dFGRS brighter than $B_{\rm MGC}=19.0$ mag.
 
The missing galaxies that are seen in the MGC galaxy catalogue can be split 
into four classes: LSBGs $(6\pm2)$ per cent; unresolved objects, $(18\pm3)$ per 
cent; blended objects, $(53\pm5)$ per cent; normal galaxies, $(19\pm3)$ per cent. 
This is in line with the findings of Pimbblet et al. (2001).

In the SDSS-EDR, there is $(1.3\pm0.1)$ per cent stellar contamination, 
$(5.3\pm1.0)$ per cent galaxies are misclassified as stars and $(0.45\pm0.07)$ 
per cent are artefacts. The SDSS-EDR galaxy catalogue is incomplete by 
$(1.8\pm0.1)$ per cent, so $(7.1\pm1.0)$ per cent of galaxies brighter than 
$B_{\rm MGC}=20.0$ mag are missing from the SDSS-EDR. The fraction of QSOs in
the stellar catalogues of the MGC and SDSS-EDR is $(2.1\pm0.4)$ per cent.

The true impact of any incompleteness on measurements of the luminosity
function can only be known with assurance by constructing a high and uniformly
complete redshift survey. We have found that even modern CCD surveys such as 
the MGC and SDSS-EDR are missing $5-7$ per cent of the galaxy sample due
to difficulties in star-galaxy separation. This means that number counts and
luminosity functions will have to be revised upwards, and the shapes may
have to be revised if the redshift distribution of these objects does not
follow the redshift distribution of the known galaxy population. Since these
galaxies are hard to separate from stars, they are likely to be compact 
galaxies, possibly from the same population as found by Drinkwater et al. 
(1999) in Fornax. They estimated that $(3.2\pm1.2)$ per cent of compact galaxies
were missed from 2dFGRS. This is compatible to our value of $(5.6\pm1.3)$ per 
cent, given that the Fornax cluster at $z=0.0046$ is significantly closer 
than the average galaxy in our sample ($z=0.1$). At the distance to Fornax 
fewer galaxies should be unresolved since they would have to have scale 
lengths $R<100$ pc. The constraints on the galaxies in our sample are
$R<2$ kpc on average, with a final fraction $(1.4\pm1.3)$ per cent of our galaxy 
sample in the same range as the Fornax cluster members. Since all these scale 
lengths are upper limits it is impossible to say for certain whether these 
constitute the same types of galaxy. These objects will be analysed in more detail 
in a later paper.

\section{Acknowledgments}

We would like to thank Keith Horne and Mike Merrifield for their comments
on NJGC's thesis which led to improvements in this paper. The MGC data were 
obtained through the Isaac Newton Group's Wide Field Camera Survey Program. 
The Isaac Newton Telescope is operated on the island of La Palma by the Isaac 
Newton Group in the Spanish Obervatorio del Roque de los Muchachos of the 
Instituto de Astrofisica de Canarias. We also thank CASU for their data 
reduction and astrometric calibration.     

Funding for the creation and distribution of the SDSS Archive has been provided by the Alfred P. Sloan Foundation, the Participating Institutions, the National Aeronautics and Space Administration, the National Science Foundation, the U.S. Department of Energy, the Japanese Monbukagakusho, and the Max Planck Society. The SDSS Web site is {\tt http://www.sdss.org/}.

    The SDSS is managed by the Astrophysical Research Consortium (ARC) for the Participating Institutions. The Participating Institutions are The University of Chicago, Fermilab, the Institute for Advanced Study, the Japan Participation Group, The Johns Hopkins University, Los Alamos National Laboratory, the Max-Planck-Institute for Astronomy (MPIA), the Max-Planck-Institute for Astrophysics (MPA), New Mexico State University, University of Pittsburgh, Princeton University, the United States Naval Observatory, and the University of Washington.

The 2dFGRS data 
were obtained via the two-degree facility on the 3.9-m Anglo-Australian 
Observatory. We would like to thank Bruce Peterson for help in accessing the 
2dFGRS database and Ivan Baldry for his useful comments on SDSS photometry. 
We thank all those involved in the smooth running and 
continued success of the 2dF and the AAO. NJGC
and DJL were funded by PPARC research studentships during the course of this 
work.

\def\bmgc{B_{\rm MGC}}
\def\btdf{b_{\rm 2dF}}
\def\bedr{B_{\rm SDSS-EDR}}
\def\bj{b_{\rm J}}
\def\rf{r_{\rm F}}
\def\japsec#1{\bigskip\noindent{\big #1}}
\def\japssec#1{\bigskip\noindent{\bf #1}}
\def\m@th{\mathsurround=0pt } 
\def\eqalign#1{\null\,\vcenter{\openup1\jot\m@th
 \ialign{\strut\hfil$\displaystyle{##}$&$\displaystyle{{}##}$\hfil
 \crcr#1\crcr}}\,}

\appendix
\section{Colour equations}

Since the present paper is concerned with a high-precision
comparison of photometry, it is necessary to take particular
care with the colour equations that relate different systems.

\subsection{MGC}

The calibration of the MGC was performed relative to Landolt
standards, which use the Johnson--Cousins system. The empirical
colour equation (with an imposed Vega zero-point) is

\begin{equation}
\bmgc = B - 0.145(B-V)
\end{equation}

\subsection{SDSS}

Here, it is necessary to distinguish clearly between (at least) four systems:

(1) USNO ($u' g' r' i' z'$)

(2) EDR ($u^* g^* r^* i^* z^*$)

(3) DR1 ($u g r i z$)

(4) AB ($u_{\rm AB} g_{\rm AB} r_{\rm AB} i_{\rm AB} z_{\rm AB}$)

\noindent
The last of these is intended to denote the ultimate SDSS system of AB
magnitudes, with the philosophy described by Fukugita et al. 1996 -- but
reflecting the fact that the filters used in reality
on the SDSS 2.5m differ slightly from the
responses assumed by Fukugita et al. The 2.5m filters also differ
significantly from the filters used by the USNO to define the network
of SDSS standard stars (Smith et al. 2002). The necessary colour equations
for the transformation are given in 
{\tt http://www.sdss.org/dr1/algorithms/ jeg\_photometric\_eq\_dr1.html}.

Here, we are mainly concerned with the cases of $g$ and $r$, for which

\begin{eqnarray}
g &= g' + 0.060[ (g'-r') - 0.53 ] \\
r &= r' + 0.035[ (r'-i') - 0.21 ]
\end{eqnarray}

An unfortunate aspect of this transformation is that the colour equation for
$r$ involves $i$ magnitudes. However, the offset $r-r'$ also 
correlates well with $g-r$, and the following equation was 
obtained, which allows us to work entirely within $(g,r)$ space:

\begin{equation}
r = r' + 0.016[ (g'-r') - 0.53 ]
\end{equation}

Using these corrections, the conversions to $BVR$ given by Smith et al. (2002)
can be cast in terms of DR1 magnitudes:

\begin{eqnarray}
B &= g + 0.39(g-r) +0.21 \\
V &= g - 0.58(g-r) -0.01 \\
R &= r - 0.15(g-r) -0.14
\end{eqnarray}

These empirical relations are very similar to the equations given by Fukugita et. al (1996):

\begin{eqnarray}
B &= g_{\rm AB} + 0.42(g_{\rm AB}-r_{\rm AB}) +0.20 \\
V &= g_{\rm AB} - 0.53(g_{\rm AB}-r_{\rm AB}) +0.00 \\
R &= r_{\rm AB} - 0.09(g_{\rm AB}-r_{\rm AB}) -0.16
\end{eqnarray}

Finally, there is the issue that the current SDSS magnitudes are not yet
believed to be zero-pointed to a perfect AB system. Blanton et al. (2003) 
give the following mean corrections, in order to obtain true AB magnitudes
from the EDR data:

\begin{eqnarray}
g_{\rm AB} &= g^* + 0.036 \\
r_{\rm AB} &= r^* + 0.015
\end{eqnarray}

The difference between the DR1 magnitudes and EDR magnitudes does not include
the correction to AB magnitudes. The difference only involves the Smith et al.
(2002) updates to the Fukugita et al. (1996) standard star system and the
improvements to the photometric pipeline detailed in Abazajian et al. (2002).
These changes are expected to produce a smaller offset than the corrections
to AB magnitudes. 

\subsection{2dFGRS}

The final calibration of the 2dFGRS for the public data release involved
recalibration of the SuperCosmos photometry, using a zero point that was 
based largely on SDSS EDR data, but with the Blanton et al. conversion 
to AB. The Fukugita et al. colour equations were assumed, and SCOS magnitudes 
were given a Vega zero point, with the following empirical colour equations:

\begin{eqnarray}
\bj &= B -0.30(B-V) \\
\rf &= R +0.16(V-R)
\end{eqnarray}

Explicitly, the relations used in terms of EDR magnitudes were:

\begin{eqnarray}
\bj &= g^* +0.130(g^*-r^*) +0.189 \\
\rf &= r^* -0.115
\end{eqnarray}

If we were to adopt the empirical conversion between $gr$ and $BVR$ given 
earlier, and ignore any small differences between EDR and DR1, these 
differences would now imply

\begin{eqnarray}
\bj &= B -0.27(B-V) +0.00 \\
\rf &= R +0.26(V-R) -0.03.
\end{eqnarray}

We thus see that the 2dFGRS $\bj$ magnitudes are very accurately on a
Vega system, and need no adjusting. Arguably the $\rf$ magnitudes are
too bright for a Vega zero point by 0.03~mag.; this is comparable to
the overall zero point uncertainty, and in any case we are not concerned
with the $\rf$ magnitudes in this paper.

\subsection{Predictions from the EDR}

In practice, we will wish to predict $\bmgc$ and $\bj$ magnitudes from the
SDSS photometry. We will ignore the small differences between the EDR and 
DR1, and use the above relations between $\bmgc$ and $\btdf$ and $(B,V)$,
relating $gr$ to $BV$ as in the revised empirical relations from Smith et 
al. This yields

\begin{equation}
\bmgc = g +0.25(g-r) +0.18 
\end{equation}

(or the identical estimator, in the case where EDR $g^*r^*$ magnitudes are
used).
For the 2dFGRS, we of course retain the colour equation used in calibration:

\begin{equation}
\btdf = g +0.13(g-r) +0.19
\end{equation}

Finally, we can eliminate SDSS data from these definitions, so that MGC 
magnitudes can be predicted from SCOS data alone:

\begin{equation}
\bmgc = b_{\rm SCOS} + 0.11(b_{\rm SCOS}-r_{\rm SCOS}) -0.04
\end{equation}

One important final point. The colour equations above match the MGC to the
SDSS-DR1 and EDR $u g r i z$, $u^*g^*r^*i^*z^*$ directly, not to the AB 
system, so using the Blanton et al. (2003) corrections to AB later is 
incorrect. The colour equations convert the $u g r i z$ to $B_{\rm MGC}$ 
whether the SDSS magnitudes are on the standard AB system or not since the 
colour equations were calculated for the $u g r i z$ filters, not the 
$u_{\rm AB}g_{\rm AB}r_{\rm AB}i_{\rm AB}z_{\rm AB}$ filters.

\end{document}